\documentclass[a4paper]{report}
\usepackage[utf8]{inputenc}
\usepackage[T1]{fontenc}
\usepackage{RJournal}
\usepackage{amsmath,amssymb,array}
\usepackage{booktabs}

\usepackage{tikz}
\usetikzlibrary{arrows,positioning,shapes.geometric}
\usetikzlibrary{calc}
\usepackage[normalem]{ulem}
\usepackage{mathtools}
\usepackage{graphicx}
\usepackage{makecell}
\usepackage{bm}

\begin{document}

\sectionhead{Contributed research article}
\volume{XX}
\volnumber{YY}
\year{20ZZ}
\month{AAAA}

\begin{article}
\title{\pkg{CovEsts}: Nonparametric Autocovariance Estimation and Analysis}
\author{by Adam Bilchouris and Andriy Olenko}

\maketitle

\abstract{
The paper introduces the R package CovEsts, which implements several nonparametric estimators for the autocovariance function. First, it presents the theoretical foundations of the implemented estimators, their properties, assumptions, and potential limitations. Next, it outlines the structure of the package and its key functions, including several methods for estimating autocovariance functions, constructing corresponding bootstrap confidence regions, and correcting the provided estimators. The package also includes diagnostic tools, such as several metrics for comparing estimators, and additional functions for broader use. The article illustrates a high degree of flexibility of the package in the selection of function parameters and the tuning of the estimators. Applications of selected estimators and package functions are illustrated using simulated data, yearly sunspot counts, and US unemployment increments data.
}

\section{Introduction}
For a stationary time series \(X(j)\), \(j = 0, 1, \dots, \) the autocovariance function is used to measure dependencies between observations separated by a lag \(h.\)
This function appears in various theoretical results and statistical applications.
For example, in the Gaussian case, the mean function and autocovariance function describe the time series entirely. Furthermore, the autocovariance function is used in the estimation of parameters for an autoregressive time series \citep[Chapter~8.1]{Brockwell1991}, model identification, and Kriging \citep[Chapter~3.2]{Cressie1993}. It is also utilised in various non-statistical fields such as digital signal processing, image analysis \citep[Chapter~3.2]{Bull2021}, and quality assessment \citep{chemo2003}, just to name a few.
Therefore, accurate estimation of the autocovariance function is an important problem in numerous statistical and data science applications.

Surprisingly, despite several other approaches in the literature, the classical standard estimator of the autocovariance function remains the predominant choice among practitioners.  Many of them are unaware of its limitations or of more accurate alternatives developed in the recent literature. This is evident in statistical software, where the majority of R and Python packages only provide a wrapper function for the classical estimator. For example, the classical autocovariance estimator realised by the \textbf{\texttt{stats::acf}} function in R,  is used by \CRANpkg{TSA} (\citeauthor{TSA_2022}, \citeyear{TSA_2022} and \citeauthor{Cryer2008}, \citeyear{Cryer2008}), \CRANpkg{astsa}~(\citeauthor{astsa_2024}, \citeyear{astsa_2024}, and \citeauthor{Shumway2025}, \citeyear{Shumway2025}), \CRANpkg{sarima} \citep{sarima_2024} and \CRANpkg{forecast} \citep{Hyndman2008}. The package \CRANpkg{forecast} also contains a tapered autocovariance estimator, \code{taperedacf()}, which is based on the results in \cite{McMurry2010, Hyndman2015}.

The motivation for the package \CRANpkg{CovEsts}~(\citeauthor{CovEsts_2025}, \citeyear{CovEsts_2025}) was that many nonparametric autocovariance estimators appeared only in research papers with no R or Python code for their computation. Also, as will be discussed in Section~\hyperref[sec:nonparametric]{2}, different estimators have different theoretical properties, which influence when they should be used. We have not found any packages dealing with potential issues regarding estimation and their corrections, apart from \CRANpkg{ncf} \citep{ncf} implementing the positive-definite adjustment from \cite{hall1994_2}. Further, no existing packages provide a unified interface and consistent data/output structure across various autocovariance estimators.

Nonparametric autocovariance estimators are particularly important due to their minimal reliance on distributional assumptions, offering greater flexibility compared to parametric approaches. Additionally, they are often used as a preliminary step in parametric modelling, where a restricted set of predefined autocovariance models is fitted to a nonparametric estimate. For example, \CRANpkg{gstat} \citep{gstat2016} does this for the (semi-)variogram. Parametric estimation was not included in \CRANpkg{CovEsts}, as this functionality for specific parametric models is already provided by existing packages.

We intend this package to be used by practitioners working with time series and spatial statistics, with applications in areas such as finance, signal processing, and climate research. We assume basic knowledge of \texttt{R} for using the functional interface of the package. However, more specialised \texttt{R} knowledge may be required for advanced inference and working with user-defined objects.

In the article, we will interchangeably use the terms random process and time series, where the former is for samples at arbitrary time locations, and the latter is used when sampled values are on a uniform sampling grid.

The paper is structured as follows. Section~\hyperref[sec:nonparametric]{2} introduces the nonparametric autocovariance estimators and some general approaches for modifying autocovariance estimators. It discusses the theoretical properties of the estimators and the drawbacks one should be aware of. Section~\hyperref[sec:overview]{3} presents the structure of the package and selected main functions. Section~\hyperref[sec:examples]{4} provides applications to three data examples: a simulated Gaussian process, yearly sunspot counts, and unemployment data, sourced from the US Bureau of Labor Statistics. Example 4 empirically studies the computational complexity of the estimators by comparing their memory and time usage.

\section{Nonparametric estimators of autocovariance functions}\label{sec:nonparametric}
There are several nonparametric autocovariance function estimators in the literature, see the reviews in \citet[Chapter~2.4]{Cressie1993}, \citet{hall1994_2}, \citet{cuevas2013study}, \citet{Durre2015} and \citet{Bilchouris_Olenko_2025}. 
This section briefly introduces the estimators that were implemented in the package and mentions some of the theoretical drawbacks to deal with when applying them.

For a weakly stationary real-valued time series \(X(j)\), \(j = 1, 2, \dots , \) the autocovariance function is defined as 
\[
\text{C}(h) \coloneq \text{E}\left[\left(X(j) - \overline{X}\right)\left(X(j + h) - \overline{X}\right)\right].
\]
and the corresponding autocorrelation at lag \(h\) is \(\rho(h):=C(h)/C(0).\)

Another related function used in the analysis of dependencies is the semivariogram. It uses the second-order moments of increments,
\[
\gamma(h) := \frac{1}{2} \text{Var}\left[X(j) - X(j + h) \right]  ,
\]
The package \CRANpkg{CovEsts} mainly focuses on estimating \(C(\cdot)\) as, under the assumption of weak stationarity, the semivariogram can be determined using the autocovariance function through the following relation
\begin{equation} \label{eqn:variogram}
    \gamma(h) = C(0) - C(h).
\end{equation}
This relation is not necessarily true when considering the estimated functions or if the semivariogram is unbounded (\citeauthor{Chils2012}, \citeyear{Chils2012} and \citeauthor{Bilchouris_Olenko_2025}, \citeyear{Bilchouris_Olenko_2025}).

Another function used in time series analysis is the partial autocorrelation function, \(\phi_{h,h}.\) Unlike the autocorrelation function, it measures the dependencies between observations \(X(t)\) and \(X(t+h)\) after removing the effects of the intermediary lags.
The partial autocorrelation function at lag $h$ can be computed using the Durbin-Levinson algorithm \citep{Durbin1960}
\begin{equation} \label{eq:durbin}
    \phi_{h, h} \coloneq \frac{\rho(h) - \sum_{k=1}^{h-1} \phi_{h - 1, k} \rho(h - j)}{1 - \sum_{k=1}^{h-1} \phi_{h - 1, k} \rho(k)}
\end{equation}
where \(\phi_{h, k} := \phi_{h - 1, k} - \phi_{h, h} \phi_{h - 1, h - k}\) for \(k = 1, 2, \dots, h - 1,\)
and \(\phi_{1, 1} := \rho(1)\). Unlike the autocorrelation function, there is no zero lag for the partial autocorrelation function. In \textbf{stats::pacf}, the formula (\ref{eq:durbin}) uses the classical autocorrelation estimator, but any autocorrelation estimate can be supplied in the package \CRANpkg{CovEsts}.

\subsection{Standard estimators}
The first two estimators implemented in \CRANpkg{CovEsts} are well-known and the most widely used. They differ only by the normalising constants:
\begin{equation} \label{eq:std_est}
\widehat{C}^{*}(h) \coloneq \frac{1}{N-h} \sum_{j=1}^{N - h} \left(X(j) - \overline{X}\right) \left(X(j+h) - \overline{X}\right)
\end{equation}
    and
\begin{equation} \label{eq:std_est_pd}
\widehat{C}^{**}(h) \coloneq \frac{1}{N} \sum_{j=1}^{N - h} \left(X(j) - \overline{X}\right) \left(X(j+h) - \overline{X}\right) ,
\end{equation}
where \(X(j)\) represents values of an observed time series, \(\overline{X}\) is the sample mean of the time series, \(N\) is the length of the observation period and \(0 \leq h \leq N-1\) is the lag for which the autocovariance function is estimated at \citep[Section~3.17]{Yaglom1987}. These estimators are typically used for equally sampled values at integer time moments or those with a constant time difference.

Even for these classical estimators, there are several issues that are not expected when applying autocovariance functions in theoretical statistical inference.
The first estimator \eqref{eq:std_est} is unbiased if the mean \(\text{E}[X(j)]\) is known and used instead of \(\overline{X},\)  but is biased otherwise \citep{Brockwell2016}. This estimator is not positive-definite. The estimator~\eqref{eq:std_est_pd} is biased but is positive-definite. This means that the second estimator gives a true autocovariance function, whilst the first only gives an estimate of a function similar to an autocovariance function. 


Another, less-known drawback is that when considering the sum over all lags for the estimated autocorrelations in \eqref{eq:std_est_pd}, the sum is always equal to \(-1/2\), regardless of the sampled values, see \cite{Hassani2009}, \citet{Hassani2012} and \cite{Bilchouris_Olenko_2025}. This causes a problem, especially when considering estimation over long time intervals or modelling long-range dependence, as the sum of estimated autocorrelation values is always equal to \(-1/2\).
Even though the estimators have desirable theoretical properties for a fixed \(h,\) when \(N  \rightarrow \infty ,\) the constant sum necessitates substantial departure of the estimates from the true autocovariance values if a large range of \(h\) is considered for a fixed \(N.\)

\subsection{Kernel regression estimators}
The next estimator, proposed in \citet{hall1994} and \citet{hall1994_2}, applies the kernel regression of pairwise autocovariances to estimate the autocovariance function,
\begin{equation} \label{eq:hall_est}
\widehat{C}_{H}(t) \coloneq \frac{\displaystyle \sum_{i=1}^{N} \sum_{j=1}^{N}  \check{X}_{ij} K\left( \left(t - \left(t_{i} - t_{j}\right)\right) / b \right) }{\displaystyle \sum_{i=1}^{N} \sum_{j=1}^{N}  K\left( \left(t - \left(t_{i} - t_{j}\right)\right) / b \right) },
\end{equation}
where \(t, t_{i}, t_{j} \in \mathbb{R},\) \(\check{X}_{ij} \coloneq \left(X(t_{i}) - \overline{X}\right) \left(X(t_{j}) - \overline{X}\right)\), \(i, j = 1, \dots, N,\) \(K(\cdot)\) is a kernel which has the properties of a symmetric probability density and \(b > 0\) is some bandwidth.
A variant of this estimator, proposed in \cite{hall1994_2}, brings the initial estimator down to zero linearly between time moments \(T_{1} > 0\) and \(T_{2} > T_{1},\)
\begin{equation} \label{eq:hall_trunc}
    \widehat{C}_{1}(t): =
	\begin{cases}
		\widehat{C}_{H}(t), & 0 \leq t \leq T_{1} \\
		\widehat{C}_{H}\left(T_{1}\right) \left(T_{2} - t\right) \left(T_{2} - T_{1}\right)^{-1}, &  T_{1} < t \leq T_{2} \\
		0 , & t > T_{2} .
	\end{cases} 
\end{equation}  
This estimator cannot be used for long-memory time series as it vanishes at \(T_{2} ,\) but is a better choice when estimating a short-range dependent autocovariance function, as the estimate will go to zero.
Unlike estimators~\eqref{eq:std_est} and \eqref{eq:std_est_pd}, these estimators can be applied for arbitrary observation grids and lags.

The estimators given by \eqref{eq:hall_est} and \eqref{eq:hall_trunc} are not necessarily positive-definite, and thus not valid autocovariance functions. Two corrections to make them positive-definite were proposed in \citet{hall1994} and \citet{hall1994_2}.

\begin{itemize}
\item[(i)] \label{make_pos_def_1}
The first correction method computes the Fourier transform of \eqref{eq:hall_est} to obtain the corresponding spectral density. Then, it makes any negative values in the spectral density equal to zero, and performs the inverse Fourier transform to obtain a positive-definite estimate of the autocovariance function.
The modification of the spectral density that corresponds to \(\widehat{C}_{H}(\cdot)\) can be expressed as \( \widetilde{\mathcal{F}}(\theta) := \max(\widehat{\mathcal{F}}(\theta), 0) \)
for every frequency \(\theta ,\) where \(\widehat{\mathcal{F}}( \cdot ) \) and \(\widetilde{\mathcal{F}} ( \cdot ) \) denote the original and modified spectral densities.

\item[(ii)]
\label{make_pos_def_2}
The second correction method considers manipulating the Fourier transform again. However, it finds the smallest frequency corresponding to a negative value in the spectral density. Then, it sets all values in the spectrum to zero whose corresponding frequencies are larger than the smallest frequency. Then, the inverse Fourier transform is taken.
The process of selecting the frequency and modifying the spectrum is as follows. Let \( \widehat{\theta} := \inf \left\{ \theta > 0 : \widehat{\mathcal{F}} \left( \theta \right) < 0 \right\}.\)  Then the spectral density is modified as follows, \( \widetilde{\mathcal{F}}(\theta) := \widehat{\mathcal{F}}(
\theta)\bm{1}\left(\theta < \widehat{\theta}\right), \) where \(\bm{1}(A)\) is the indicator function of a set \(A.\)
The drawback is that this correction can fail to produce a meaningful result if \(\widehat{\theta}\) is a small frequency, as the modified estimator of the autocovariance function consists of a sum of only a few cosines.
Both of these correction methods can be applied to any estimator of the autocovariance function, one is not restricted to just estimators~\eqref{eq:hall_est} and \eqref{eq:hall_trunc}.

\end{itemize}

\subsection{Tapered estimators}
The following estimator, proposed in \cite{Dahlhaus1987}, takes the edge effect into account.
The edge effect reflects the situation that points closer to the boundaries of an observation region can have neighbouring observations that are outside of the study region. Thus, some dependencies may not be properly reflected, which can introduce bias. To deal with this, the estimator assigns weights for each location depending on how close it is to the boundary, where lower weights are given to locations closer to the boundaries:
\begin{equation} \label{eq:tapered_est}
    	\widehat{C}^{a}_{N}(h) \coloneq \frac{\displaystyle \sum_{j = 1}^{N - h} \left( X(j) - \overline{X}\right) \left( X(j + h) - \overline{X} \right) a\left( (j - 1/2)/N; \rho \right) \; a \left( (j + h - 1/2)/N; \rho \right) }
        {\displaystyle H_{2, N}(0) },    
\end{equation}
where the normalising factor is
\[
H_{2, N}(0) := \sum_{s=1}^{N} a ( (s - 1/2)/N; \rho )^{2} ,
\] \(a(\cdot; \cdot)\) is a taper function on the interval \([0, 1]\) with the smoothness parameter \(\rho \in (0, 1],\)
\[
a(u; \rho) := 
\begin{cases}
		w(2u/\rho) ,& 0 \leq u < \frac{1}{2}\rho, \\
		1 ,& \frac{1}{2} \rho \leq u \leq \frac{1}{2}, \\
		a(1-u; \rho) ,& \frac{1}{2} < u \leq 1,
\end{cases} 
\]
and \(w(\cdot)\) is a continuous nondecreasing function on \([0, 1]\) with \(w(0)=0\) and \(w(1)=1.\)
We will refer to \(w(\cdot)\) as a window function.
The estimator~\eqref{eq:tapered_est} is positive-definite and biased, but the bias is negligible asymptotically. This estimator is applied to observations on an integer grid.

\subsection{Splines estimators}
Unlike the other estimators, the following estimator does not use observations directly. It is based on an approximation of \(\widehat{C}(\cdot)\) by completely monotone basis functions proposed in \cite{Choi2013} \\[-2mm]
\begin{equation} \label{eq:splines_est}
    \widehat{C}^{B}(h) \coloneq \sum_{j=1}^{m + p} \beta_{j} f_{j}^{(p - 1)} \left(h^{2}\right) ,
\end{equation}
where \(\beta_{j} \geq 0\), \(f_{j}^{(l)} (x) \coloneq \int_{0}^{1} (m + 1) t^{x} B_{j + 1}^{(l)} (t) \text{d} t, \) \(B_{j}^{(l)}(\cdot)\) is the \(j^{\text{th}}\) B-spline of order \(l\) and and \(j = 1, 2, \dots , m + p \).
The constants \(\beta_{j}\) are chosen via weighted least squares, where the objective function is
\[
\sum_{i = 1}^{L} w_{i} \left( \widehat{C}(h_{i}) - \sum_{j=1}^{m + p} \beta_{j} f_{j}^{(p - 1)} \left(h_{i}^{2}\right) \right)^{2} ,
\]
\(\widehat{C}(\cdot)\) is an autocovariance function estimator, \(\{ h_{1} , \dots , h_{L} \}\) is a set of lags and \(\{ w_{1}, \dots , w_{L} \}\) is a set of weights. \cite{Choi2013} uses \eqref{eq:std_est} as \(\widehat{C}(\cdot)\), however, one can use any autocovariance estimator. For the choice of weights \(w_{i} = \left(N-h_{i}\right) / \left(1 - \widehat{C}(h_{i})\right)^{2} \) was proposed in \cite{Cressie1985}.

This estimator can calculate the estimated autocovariance at an arbitrary lag once the fitting process is done. However, the lags used during the fitting process should be chosen such that they are the same as in \(\widehat{C}(\cdot).\)
As estimator~\eqref{eq:splines_est} is constructed using completely monotone basis functions, and \(\beta_{j} \geq 0,\) it is nonnegative. This estimator is positive-definite function. Further, it is also bounded from below by zero, meaning this estimator cannot be used when the autocovariance is below zero, and due to the monotonicity, it cannot be used if cyclicality is present.

\subsection{Kernel correction estimators}
The next estimator is simply a modification of any estimator. It was proposed to remove estimation \dfn{wave} artefacts \citep[Section~3.17]{Yaglom1987}. The waves are present in various estimated autocovariance functions and are more prominent as the estimation lag increases, as fewer points are available to compute the autocovariance function. This estimator also helps to reduce constant summation effects, which were discussed earlier for estimators~\eqref{eq:std_est} and \eqref{eq:std_est_pd}. It is defined as
\begin{equation} \label{eqn:kernel_correction}
    \widehat{C}_{T}^{(a)}(h) \coloneq a_{T}(h) \widehat{C}(h) ,
\end{equation}
where \(a_{T}(\cdot)\) is a kernel function, which approaches zero as \(\left| h \right|\) increases and \(\widehat{C}(\cdot)\) is any autocovariance estimator \citep[Section~3.17]{Yaglom1987}.
Typically, \(a_{T}(h) \coloneq a(h / N_{T})\), where \(N_{T}\) is some constant, usually 10\% of the number of observations. 
If \(a_{T}(\cdot)\) is chosen as positive-definite and \(\widehat{C}(\cdot)\) is positive-definite, then \(\widehat{C}_{T}^{(a)}\) will also be positive-definite.

A combination of \eqref{eq:std_est_pd} and \eqref{eqn:kernel_correction} can also be used to reduce the waves in the estimate and guarantee a finite range of dependencies:
\begin{equation} \label{eqn:kernel_correction_std_pd}
    \widehat{C}_{T}^{**}(h) = a_{T}(h) \widehat{C}^{**}(h) .
\end{equation}

\subsection{Linear shrinking}
A linear shrinkage correction method, introduced by \citet{Devlin1975}, adjusts the estimated autocorrelation matrix \(\bm{R}\) through the following transformation
\begin{equation} \label{eqn:linear_shrinking}
    \widetilde{\bm{R}} \coloneq \lambda \bm{R} + (1 - \lambda)\bm{I}_{p} ,
\end{equation}
where \(\widetilde{\bm{R}}\) is the shrunken autocorrelation matrix, \(\lambda \in [0, 1]\) is the shrinking coefficient and \(\bm{I}_{p}\) is the \(p \times p\) identity matrix, often called the shrinkage target \citep{Rousseeuw1993}.
\(\lambda\) is chosen as the maximal value for which \(\widetilde{\bm{R}}\)
remains positive-definite.

\subsection{Block bootstrap}

The moving block bootstrap, independently developed by \cite{Kunsch1989} and \cite{Liu1992}, allows for the resampling of time series data with dependencies without relying on any parametric assumptions~\citep[Chapter~2.5]{Lahiri2003}.
For a time series \(X(1), \dots , X(n)\), first,  construct \(n - \ell + 1\) blocks of length \(\ell,\)
\(\mathcal{B}_{i} = \left( X(i), \dots , X(i + \ell - 1) \right), \) for \(i = 1, \dots , n - \ell + 1.\)
Then, the blocks are sampled in the following way.
Let \(I_{1}, \dots, I_{k}\) be \(k\) independent and identically sample values, from the discrete uniform distribution on \(\left\{ 1,\dots, n - \ell + 1 \right\},\) which are the block indices, that is \(\mathcal{B}_{I_{i}}.\)
To construct a bootstrapped time series, join the randomly sampled blocks \(\mathcal{B}_{I_{1}}^{*}, \dots, \mathcal{B}_{I_{k}}^{*},\) resulting in the time series \(X^{*}(1), \dots, X^{*}(k\ell),\) where \(*\) denotes the sampled versions of the blocks and time series \cite[Chapter~2.5]{Lahiri2003}. If $k\ell>n,$ the moving block sampled time series is truncated at $n.$

The moving block bootstrap suffers from a boundary effect, where lesser weights are given to observations at the beginning and end of the sampled time series \citep[Chapter~2.7]{Lahiri2003}. 
A modified method to construct bootstrap samples is the circular bootstrap, proposed by \cite{Politis1992}, which addresses this issue.
Instead of the time series \(X(1) , \dots , X(n)\) being observed on the line, it is considered to be observed on the circle. This results in the observation \(X(n + k)\) being the same as \(X(k)\) for \(k = 1, \dots , n.\)
 For \(i = 1, \dots , n ,\) blocks are constructed in a similar, but circular fashion. For example, the block \(\mathcal{B}_{n - \ell + 2} = \left(X(n - \ell + 2) , \dots , X(n), X(n + 1) \right)\) is the same as \(\left(X(n - \ell + 2) , \dots , X(n), X(1) \right) .\)
The procedure to construct a circular bootstrap time series is the same as for the moving block bootstrap.
However,  the block indices are instead sampled from \(\{1, \dots , n\}\).

The block length parameter choice is flexible, although some guidelines exist, see \cite{Hall1995}, \cite{Politis2004} and \cite{Arteche2024}. For example, a block length \(\ell \sim C n^{1/k}, k = 3, 4, 5,\) is recommended in \citep{Hall1995}.
For a time series with long memory, the block bootstrap may give inconsistent results, see \cite{Lahiri1993}.
In such cases, a larger block size should be chosen to capture the dependency structure.

In the context of autocovariance estimation in this paper, first, a block bootstrap method is used to sample multiple time series. Then, a specified autocovariance estimation method is applied to compute the estimated autocovariance for each sampled time series. The bootstrap estimate is the average of all individual estimators, and the corresponding bootstrap confidence region is constructed from the bootstrap confidence intervals at each lag $t$ of the considered autocovariance estimate. 

Table~\ref{tab:tab_summary} summarises some properties of the estimators presented in this section, including their input requirements, assumptions of grid regularity, positive-definiteness, computational efficiency, suitability for long-memory processes, and disadvantages that applied users should be aware of.

Additional theoretical results and formulas will be introduced later when needed.
{
\setlength{\tabcolsep}{3pt}   
\renewcommand{\arraystretch}{1.3}
\begin{widetable}[hbt!]
    \centering
 {
    \begin{tabular}{l c c c c c c} \toprule
{Estimator} &
{Input} &
{Grid} &
{P.d.} &
\makecell{Computational\\ efficiency} &
\makecell{Suitable for\\ long memory} &
{Disadvantages} \\
\midrule
        Standard~(3) & Data & Regular & No & Fast & No &
        \makecell{Constant Cov.est.\\ summation}\\ \midrule

        Standard~(4) & Data & Regular & Yes & Fast & No &  \makecell{Constant Cov.est.\\ summation}  \\ \midrule

        Hall's~(5) & Data & Both & No & Slow/unstable & Yes & \makecell{Slow for long\\ time series}  \\ \midrule

        Hall's Truncated~(6) & Data & Both & No & Slow/unstable & No & \makecell{Slow for long\\ time series} \\ \midrule

        Hall's Correction~(i) & Cov.est. & Regular & Yes & Fast & No & \makecell{Inflates est.variance} \\ \midrule

        Hall's Correction~(ii) & Cov.est. & Regular & Yes & Fast & No &  \makecell{May degenerate\\ to few cosines} \\ \midrule

        Tapered~(7) & Data & Regular & Yes & Fast & Yes & \makecell{Tapering may\\ cause bias} \\ \midrule

        Splines~(8) & \makecell{Data \&\\ Cov.est.}    & From Cov.est. & Yes & Slow/unstable & No &  \makecell{Cov.est. is always\\ nonnegative} \\ \midrule

        Kernel Correction~(9) & Cov.est. & Regular & No & Fast & No & \makecell{Removes long\\ memory nature} \\ \midrule

        Kernel Correction~(10) & Cov.est. & Regular & No & Fast & No & \makecell{Removes long\\ memory nature}  \\ \midrule

        Shrinkage Correction~(11) & Data  & Regular & No & From Cov.est. & \makecell{Target\\ dependent} & \makecell{High bias if\\ wrong target} \\ \midrule

        \makecell[l]{Moving block\\ bootstrap average} & Data  & From Cov.est. & No & From Cov.est. & From Cov.est. & \makecell{Weighting\\  boundary effect} \\ \midrule

        \makecell[l]{Circular bootstrap\\ average} & Data & From Cov.est. & No & From Cov.est. & From Cov.est. & \makecell{Duplicated\\ information, \\ Edge jumps} \\ \bottomrule
    \end{tabular}}
    \caption{Summary of selected properties of the estimators}
    \label{tab:tab_summary}
\end{widetable}
}
    
\section{Package structure and function overview} \label{sec:overview}

This section provides a high-level overview of the package, including the types of functions and their relations, function structures, and the parameters of selected functions.

\subsection{Package interface and S3 objects overview}

The package \CRANpkg{CovEsts} consists of several functions to compute the autocovariance estimators discussed in the previous section.
For the \CRANpkg{CovEsts} package, we deliberately aimed to keep the number of package dependencies minimal, so it only uses packages that come with base R, \pkg{stats} for \texttt{acf}, \texttt{optim} and \texttt{fft}, \pkg{graphics} for \texttt{legend}, \texttt{lines} and \texttt{polygon}, and \pkg{parallel} for \texttt{parLapply}, \texttt{makePSOCKcluster}, \texttt{clusterExport} and \texttt{stopCluster}. This was done to make the package self-contained.

\begin{figure}[htb!]
\centering
\begin{tikzpicture}
    \tikzstyle{block} = {draw, rectangle, align=center, minimum width=2cm, minimum height=1cm}
    \tikzstyle{line} = [draw, -latex']

    \node [block,align=center]  (cov_ests) {\uline{Autocovariance Estimators}\\
        {\ttfamily standard\_est}\\
        {\ttfamily adjusted\_est}\\
        {\ttfamily truncated\_est}\\
        {\ttfamily tapered\_est}\\
        {\ttfamily splines\_est}\\
        {\ttfamily kernel\_est}\\
        {\ttfamily corrected\_est}\\
        {\ttfamily to\_vario}\\
        {\ttfamily to\_pacf}\\
        {\ttfamily block\_bootstrap}};

    \node [block, align=center, left = 1.2cm of cov_ests, yshift=1.3cm] (general) {\uline{General Functions}\\
        {\ttfamily dct\_1d}\\
        {\ttfamily idct\_1d}\\
        {\ttfamily make\_pd}\\
        {\ttfamily nearest\_pd}\\
        {\ttfamily shrinking}};


      \node [block, align=center, below = 0.4cm of general] (kernels) {\uline{\shortstack{Kernel and Window\\Functions}}\\
        {\ttfamily kernel\_ec}\\
        {\ttfamily kernel\_symm\_ec}\\
        {\ttfamily window\_ec}\\
        {\ttfamily window\_symm\_ec}};
        
  \node [block, align=center, right = 1.2cm  of cov_ests, yshift=1.1cm] (metrics) {\uline{Metric Functions}\\
        {\ttfamily area\_between}\\
        {\ttfamily max\_distance}\\ 
        {\ttfamily spectral\_norm}\\
        {\ttfamily check\_pd}\\
        {\ttfamily hilbert\_schmidt}\\
        {\ttfamily mse}};

 \node [block, align=center,  below = 0.4cm of metrics] (helpers) {\uline{Helper Functions}\\
        {\ttfamily taper}\\
        {\ttfamily normalise\_acf}\\
        {\ttfamily bootstrap\_sample}\\
        {\ttfamily starting\_locs}};

   \draw [<-] ($(cov_ests.north west)!0.43!(cov_ests.south west)$) -- (general.east);
\draw [<-] ($(cov_ests.north west)!0.7!(cov_ests.south west)$) -- (kernels.east);
 \draw [->] ($(cov_ests.north east)!0.43!(cov_ests.south east)$)(cov_ests) -- (metrics.west);
    \draw [<-] ($(cov_ests.north east)!0.7!(cov_ests.south east)$) -- (helpers.west);

\end{tikzpicture}
\caption{Types and links between main package functions.}
\label{fig:functions_group}
\end{figure}
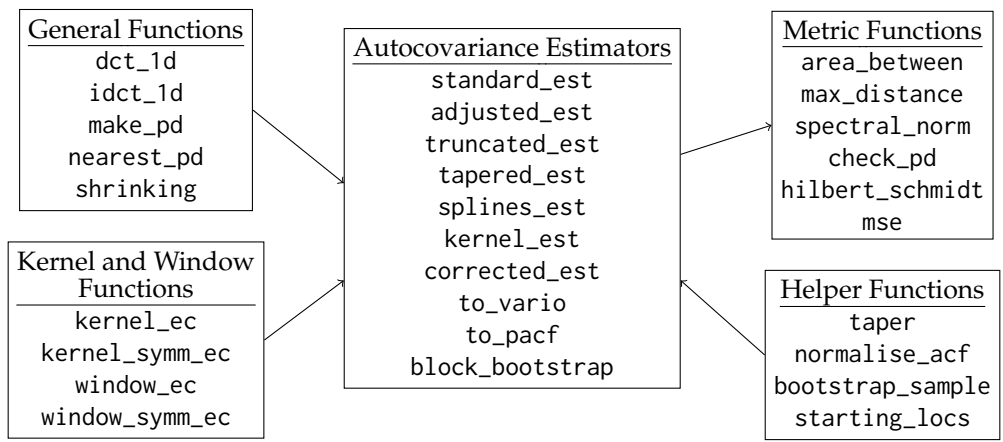

Figure~\ref{fig:functions_group} groups the main functions depending on their functionality and relations to the autocovariance estimation functions.
The main group, \textit{Autocovariance Estimators}, consists of the autocovariance function estimators discussed in Section~\hyperref[sec:nonparametric]{2}, whilst the remaining groups are used for intermediary and complementary calculations, except those that compare autocovariance estimates. Note that this figure does not include all functions in the package. In particular, non-user-facing functions are omitted.

In the main group, there are nine functions that estimate the autocovariance function or its modifications.
As the parameters are often shared or similar between the estimator functions, to avoid repetition, only one set of parameters is shown in Table~\ref{tab:truncated_est_params}. As can be seen from the table, the functions are vectorised.
\renewcommand{\arraystretch}{1}
\begin{table}[htb!]
    \centering
    \begin{tabular}{l p{0.77\linewidth}}\toprule
        Parameter & Explanation \\ \midrule
        \texttt{X} & A vector representing observed values of the time series. \\
        \texttt{x} & A vector of lags. \\
        \texttt{t} & The arguments at which the autocovariance function is calculated at. \\
        \texttt{T1} & The first truncation point, \(T_{1} > 0.\) \\
        \texttt{T2} & The second truncation point, \(T_{2} > T_{1} > 0.\) \\
        \texttt{b} & Bandwidth parameter, greater than 0. \\
        \texttt{kernel\_name} & The name of the symmetric kernel function to be used. Possible values are: gaussian, wave, rational\_quadratic, and bessel\_j. Alternatively, a custom kernel function can be provided. \\
        \texttt{kernel\_params} &  A vector of parameters of the kernel function. \\
        \texttt{custom\_kernel} & If a custom kernel is to be used or not. Defaults to \texttt{FALSE}. \\
        \texttt{pd} & Whether a positive-definite estimate should be used. Defaults to \texttt{TRUE}. \\
        \texttt{type} & Compute either the `autocovariance' or `autocorrelation'. Defaults to `autocovariance'. \\
        \texttt{meanX} & The average value of \texttt{X}. Defaults to \texttt{mean(X)}. \\ 
        \texttt{parallel} & Whether or not the computations should be done in parallel or not. Defaults to \texttt{FALSE}. \\
        \texttt{ncores} & The number of cores to be used in the parallel computations. Defaults to the number cores - 1 (threads if hyperthreading is available), calculated from \texttt{parallel::detectCores() - 1}. \\
        \texttt{cl\_export} & A vector of any additional functions or variables to export for parallel computations. This may be required if \code{estimator} is not within the package. Defaults to \texttt{NULL}. \\
        \texttt{cl} & An optional cluster object created by \code{parallel::makeCluster}. Defaults to \code{NULL}, which creates a temporary PSOCK cluster. \\ \midrule
        Return & A CovEsts S3 object containing the truncated kernel regression estimates, the lags, the estimated type and the estimator used. \\ \bottomrule
    \end{tabular}
    \caption{Example of parameters for \texttt{truncated\_est}}
    \label{tab:truncated_est_params}
\end{table}

The estimator functions are called through a single function, accepting a time series as the first argument, and additional arguments depending on the estimation method. As the first argument is the time series, the pipe operator \code{|>} can be used for all estimators, for example, \code{X\,|>\,truncated\_est(other arguments)}.
All autocovariance estimator functions return a \textbf{\texttt{CovEsts}} S3 object. These S3 objects are lists containing four elements,
\begin{itemize}
    \item \code{acf}: the estimated autocovariance/autocorrelation/partial autocorrelation values,
    \item \code{lags}: the lag indices used to compute the estimates on,
    \item \code{est\_type}: the type of estimate, namely `autocorrelation', `autocovariance' or `partial',
    \item \code{est\_used}: the estimator function used.
\end{itemize}

In addition to the \textbf{\texttt{CovEsts}} S3 objects, there are also the \textbf{\texttt{VarioEsts}} and \textbf{\texttt{BootEsts}} S3 objects for variograms computed using \texttt{to\_vario} and for block bootstrap estimates.
The \textbf{\texttt{VarioEsts}} object has the following elements,
\begin{itemize}
    \item \code{acf}: the estimated variogram values,
    \item \code{lags}: the lag indices used to compute the estimates on,
    \item \code{est\_type}: `to\_vario'.
\end{itemize}

The \textbf{\texttt{BootEsts}} has several more elements, 
\begin{itemize}
    \item \code{avg\_acf}: the average bootstrap autocovariance/autocorrelation function estimate,
    \item \code{lags}: the lag indices used to compute the estimates on,
    \item \code{acf\_orig}: the nonbootstrapped autocovariance/autocorrelation estimate,
    \item \code{acf\_mat}: a matrix of all of the bootstrap estimates,
    \item \code{conf\_lower}: the lower bounds for the estimated pointwise confidence interval,
    \item \code{conf\_upper}: the upper bounds for the estimated pointwise confidence interval,
    \item \code{est\_type}: the type of estimate, namely `autocorrelation', `autocovariance',
    \item \code{est\_used}: the estimator function used,
    \item \code{boot\_type} is either `moving' or `circular' depending on the type of block bootstrap used,
    \item \code{alpha} is the \(\alpha\) value used to compute the confidence intervals.
\end{itemize}

All these S3 objects have \texttt{print}, \texttt{plot} and \texttt{lines} methods. The \texttt{lines} method allows easy comparison between the obtained estimates. For \textbf{\texttt{CovEsts}} and \textbf{\texttt{VarioEsts}} it overlays a line directly on the plot. The \code{lines} method for \textbf{\texttt{BootEsts}} plots a single averaged estimate across all bootstrap samples, rather than displaying the individual estimates corresponding to each bootstrap sample.

S3 objects were chosen over S4 and R6 due to their simplicity, both internally and in terms of user-facing flexibility. Namely, they allow plotting and printing the obtained estimates as ordinary \texttt{R} objects, without requiring inspection of the underlying list structure. This is particularly useful when plotting multiple results together, enabling an easy comparison. Further, if a user wishes to extract the estimated values or any other elements, this can be done straightforwardly.

\subsection{Main estimators}
To compute the standard estimator~\eqref{eq:std_est}, one uses the function
\begin{example}
standard_est(X, pd = TRUE, maxLag = length(X) - 1, x = 0:length(X),
        type = c("autocovariance", "autocorrelation"), meanX = mean(X)) ,
\end{example}
where \texttt{maxLag} \(\leq N - 1\) is the maximum lag for which the estimated autocovariance function is computed and \texttt{x} are the indices for which the time series was observed on. To obtain the estimate~\eqref{eq:std_est_pd}, \texttt{pd = TRUE} should be used instead.
As can be seen, this function assumes the basic case of equally spaced observations on a consecutive grid of points, such as the integers or those with a constant difference of the observation period.
This function, along with all other estimator functions with the "\code{\_est}" suffix, returns a \textbf{\texttt{CovEsts}} S3 object.

To provide examples of other function calls, we consider estimators~\eqref{eq:hall_est} and \eqref{eq:hall_trunc} that can be computed through the commands
\begin{example}
adjusted_est(X, x, t, b,
        kernel_name = c("gaussian", "wave", "rational_quadratic", "bessel_j"),
        kernel_params=c(), pd = TRUE, type = c("autocovariance", "autocorrelation"),
        meanX = mean(X), custom_kernel = FALSE, parallel = FALSE,
        ncores = parallel::detectCores() - 1, cl_export = NULL, cl = NULL) 

truncated_est(X, x, t, T1, T2, b,
        kernel_name = c("gaussian", "wave", "rational_quadratic", "bessel_j"),
        kernel_params = c(), pd = TRUE, type = c("autocovariance", autocorrelation"),
        meanX = mean(X), custom_kernel = FALSE, parallel = FALSE,
        ncores = parallel::detectCores() - 1, cl_export = NULL, cl = NULL) .
\end{example}
These estimators can use sampled values on an arbitrary grid and set of lags.
For the parameters for these two estimators, refer to Table~\ref{tab:truncated_est_params}.

As several estimators include options for kernel smoothing and adjustment, a list of possible symmetric kernels can be found in the discussion of the kernels and window functions later in this section and in Table~\ref{tab:kernels}.
For the considered estimators, if a custom kernel is chosen, it must have the properties of a symmetric probability density.

These two functions also allow for parallelism. The other estimators either do not support parallelisation or do not benefit from it, as the call-based R functions are sufficiently fast. Parallelisation is done over the argument \code{t}, which represents the lags. The user can pass their own cluster, otherwise, a temporary PSOCK cluster is created. This can be useful if the user is on an operating system that supports \texttt{fork()}, allowing a cluster created using \code{parallel::makeForkCluster} to be used instead.

Another example, the splines estimator, defined by \eqref{eq:splines_est}, is called as
\begin{example}
splines_est(X, x, estCov, p, m, maxLag = length(X) - 1,
        type = c("autocovariance", "autocorrelation"), initial_pars = c(),
        control = list('maxit' = 1000)) ,
\end{example}
where \texttt{estCov} is an estimated autocovariance function used during the fitting process, which can either be a numeric vector or a \textbf{\texttt{CovEsts}} S3 object generated by one of the other autocovariance function estimators. The parameter \texttt{p} is the order of the splines, \texttt{m} is the number of nonboundary knots, \texttt{initial\_pars} and \texttt{control} are optional parameters used during optimisation process (see \textbf{\texttt{stats::optim}} for \texttt{control}), where \texttt{initial\_pars} is an \(\texttt{m} + \texttt{p}\) vector whose default values are 0.5. This estimator can fail to produce an output if neither optimisation algorithm converges. The optimisation algorithms used are Nelder-Mead and L-BFGS-B.

\begin{figure}[htbp]
    \centering
    \begin{tikzpicture}
        \tikzstyle{block} = {draw, rectangle, align=center, minimum width=2cm,minimum height=1cm}
        \tikzstyle{line} = [draw, -latex'],
        \node [block,align=center]  (spline_est) {{\ttfamily spline\_est}};
    
        \node [block,align=center , below = 1cm of spline_est]  (get_taus) {{\ttfamily get\_all\_tau}};
        \node [block,align=center , below = 1cm of get_taus]  (generate_knots) {{\ttfamily generate\_knots}};
    
        \node [block,align=center , right = of get_taus]  (splines_df) {{\ttfamily get\_splines\_df}};
        \node [block,align=center , below = 1cm of splines_df] (adjusted_spline) {{\ttfamily f\_j\_l}};
    
        \node [block,align=center , right = 5cm of spline_est]  (optim1) {{\ttfamily optim}};
        \node [block,align=center , right = 8cm of spline_est]  (optim2) {{\ttfamily optim}};
    
        \node [block,align=center , below = 1cm of optim1]  (solve_spline1) {{\ttfamily solve\_spline}};
        \node [block,align=center , below = 1cm of optim2]  (solve_spline2) {{\ttfamily solve\_spline}};
        
        \draw [->] (spline_est) -- (get_taus) node[midway, left] {1};
        \draw [->] (get_taus) -- (generate_knots) node[midway, left] {1.1};
    
        \draw [->] (spline_est) -| (splines_df) node[midway, below left] {2};
        \draw [->] (splines_df) -- (adjusted_spline) node[midway, left] {2.1};
        \draw [->] (adjusted_spline) to [out=270, in=0, looseness=5] node[right] {2.1.1} (adjusted_spline);
        
        \draw [->] (spline_est) to [out=5, in=175] node[below right] {3} (optim1);
        \draw [->] (spline_est) to [out=10, in=170] node[above right] {3} (optim2);
        \draw [->] (optim1) -- (solve_spline1) node[midway, left] {3.1};
        \draw [->] (optim2) -- (solve_spline2) node[midway, left] {3.1};
    \end{tikzpicture}
    \caption{Example of a high-level structure of \texttt{spline\_est} and the functions it calls.}
    \label{fig:splines_est}
\end{figure}
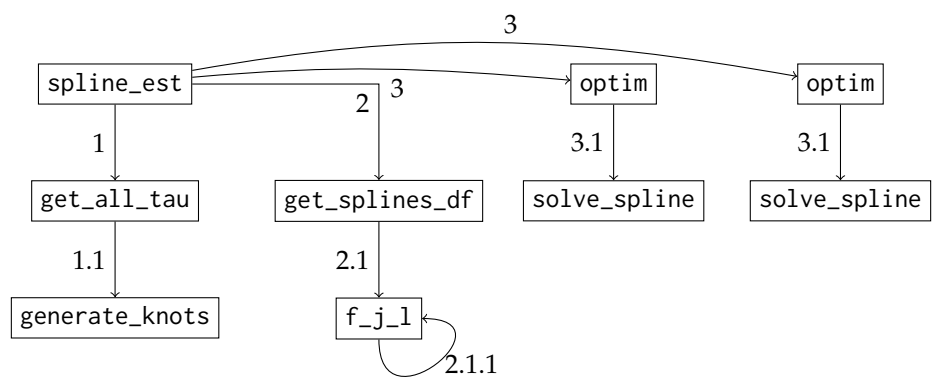

Many of the estimators considered in this package require non-trivial multistep calculations. As an example, the high-level structure of the function \texttt{splines\_est} and the steps required to compute it are shown in Figure~\ref{fig:splines_est}.
Arrows represent function calls to other functions for obtaining values from them. The function calls for \texttt{optim} are repeated as it is called twice using different optimisation algorithms, where the one with the lowest error is selected.
A whole number placed near an arrow indicates the corresponding step of the computation in which the function operates. The number after the whole number, is the order in which a function is called by the level preceding it. 

The semivariogram estimate can be computed from an autocovariance function estimate \texttt{estCov}, either a numeric vector or a \textbf{\texttt{CovEsts}} S3 object, as per \eqref{eqn:variogram}, using the command 
\begin{example}
to_vario(estCov) .
\end{example}

The partial autocorrelation for any estimated autocovariance or autocorrelation function \texttt{estCov}, either a numeric vector or a \textbf{\texttt{CovEsts}} S3 object, can be computed using
\begin{example}
to_pacf(estCov) .
\end{example}

Block bootstrap can be performed using the following command
\begin{example}
block_bootstrap(X, maxLag, x = 0:length(X), n_bootstrap = 100,
        l = ceiling(length(X)^(1/3)), estimator = standard_est,
        type = c("autocovariance", "autocorrelation"), alpha = 0.05,
        boot_type = c("moving", "circular"), parallel = FALSE,
        ncores = parallel::detectCores() - 1, cl_export = NULL,
        boot_seed = NULL, cl = NULL, ...) ,
\end{example}
where \code{n\_bootstrap} is the number of times to run block bootstrap, \code{l} is the block length, \code{estimator} is the estimator function one wishes to use, where estimator~\eqref{eq:std_est_pd} is chosen by default, \code{alpha} is the level of significance for the pointwise confidence intervals, \code{boot\_type} is either \code{'moving'} or \code{'circular'}. This function supports parallelism, see Table~\ref{tab:truncated_est_params} for the relevant parameters. "\code{...}" are any other parameters that are to be passed into \code{estimator}. The parallelism is performed over \code{n\_bootstrap} iterations.

\subsection{Kernels, windows and associated estimators}
Several standard kernels,  symmetric kernels and window functions available in the literature are realised in the package. We also provided the option for user-defined kernels, symmetric kernels, window functions and symmetric window functions. Tables~\ref{tab:kernels} and \ref{tab:windows} provide lists of the main available kernels and window functions in the package. The symmetric kernels and symmetric window functions will be mentioned only briefly, as they are modifications of the main kernels.
\begin{table}[htbp!]
    \centering
    \begin{tabular}{l l l} \toprule
        Kernel Name                    &  Equation \(a(x; \theta, \nu, d, \alpha, \beta)\)   & Constraints                 \\ \midrule
        \texttt{gaussian}            & \(\exp(-x^{2} / \theta)\)   &                             \\
        \texttt{exponential}         & \(\exp(-x / \theta)\)       &                             \\
        \texttt{wave}                & \(\begin{cases} \frac{\theta}{x} \sin(x / \theta), & x \neq 0 \\ 0, & x = 0 \end{cases} \) & \\
        \texttt{rational\_quadratic} & \(1 - {x^{2}}/{(x^{2} + \theta)} \) & \\
        \texttt{spherical}           & \(\begin{cases} 1 - {3x}/{(2\theta)} + \left( {x}/{\theta} \right)^{3}/2, & x < \theta \\ 0, & \text{otherwise} \end{cases}\) & \\
        \texttt{circular}            & \(\begin{cases} \frac{2}{\pi}\arccos\left( {x}/{\theta} \right) - \frac{2x}{\pi\theta} \sqrt{ 1 - \left( {x}/{\theta} \right)^{2} }, & \hspace{-2mm} x < \theta \\ 0, & \hspace{-9mm}\text{otherwise} \end{cases}\) & \\
        \texttt{matern}              & \(\left(\sqrt{2\nu} {x}/{\theta}\right)^{\nu} \left(2^{\nu - 1}  \Gamma(\nu)\right)^{-1} K_{\nu}(\sqrt{2\nu} {x}/{\theta})\) & \(\nu > 0\) \\
        \texttt{bessel\_j}           & \(2^{\nu} \Gamma(\nu + 1) J_{\nu}(x / \theta) (x / \theta)^{-\nu}\) & \(\nu \geq d/2  -1\) \\
         \texttt{cauchy}             & \((1 + (x / \theta)^{\alpha})^{-(\beta / \alpha)}\) & \(\alpha \in (0, 2], \beta \geq 0\) \\ \bottomrule 
    \end{tabular}
    \caption{List of main kernels}
    \label{tab:kernels}
\end{table}

The kernel and window functions are called through
\begin{example}
kernel_ec(x, name = c("gaussian", "exponential", "wave", "rational_quadratic",
        "spherical", "circular", "bessel_j", "matern", "cauchy"), params=c(1)) ,
kernel_symm_ec(x, name = c("gaussian", "wave", "rational_quadratic", "bessel_j"),
        params=c(1)) ,
window_ec(x, name = c("tukey", "triangular", "sine", "power_sine", "blackman",
        "hann_poisson", "welch"), params=c(1)) ,
window_symm_ec(x, name = c("tukey", "triangular", "sine", "power_sine", "blackman",
        "hann_poisson", "welch"), params=c(1)) ,
\end{example}
where the parameters are given in the formulas in Tables~\ref{tab:kernels} and \ref{tab:windows}.

The symmetric kernels (\texttt{gaussian}, \texttt{wave}, \texttt{rational\_quadratic} and \texttt{bessel\_j}) are symmetric versions of the kernels in Table~\ref{tab:kernels}, but they are standardised so that their area is 1.
For the symmetric window functions, all options are the same as the window functions, however, they are defined as \(1 - w(\left|x\right|;a)\) for \(x \in [-1, 1]\) and \(0\) elsewhere, where \(w(x;a)\) is a window function.
For \texttt{kernel\_ec}, its argument \texttt{x} is nonnegative, for \texttt{kernel\_symm\_ec} and \texttt{window\_symm\_ec}, their arguments \texttt{x} can be negative, and for \texttt{window\_ec}, it must be within 0 and 1.
\begin{table}[htb!]
    \centering
    \begin{tabular}{l l l} \toprule
        Kernel Name             &  Equation \(w(x; a)\)      & Constraints                 \\ \midrule 
        \texttt{tukey}        & \(\frac{1}{2} - \frac{1}{2} \cos(\pi x)\)   &   \\
        \texttt{triangular}   & \(x\) &   \\
        \texttt{sine}         & \(\sin(\pi x / 2)\) &   \\
        \texttt{power\_sine}   & \(\sin^{a}(\pi x / 2)\) & \(   a > 0\) \\
        \texttt{blackman}     & \(( (1 - a) / 2) - \frac{1}{2} \cos(\pi x) + \frac{a}{2} \cos(2 \pi x)\)   & \(   \left|a\right| \leq 0.25 \) \\
        \texttt{hann\_poisson} & \((1/2) (1 - \cos(\pi x)) \exp( - (a \left|1 - x \right|) )\)   & \( a \in \mathbb{R}\) \\
        \texttt{welch}        & \(1 - (x - 1)^2 \)   &   \\ \bottomrule
    \end{tabular}
    \caption{List of main window functions, where \(x \in [0, 1]\)}
    \label{tab:windows}
\end{table}

The package offers several functions to deal with the potential issues in the estimated autocovariance functions, which were discussed in Section~\hyperref[sec:nonparametric]{2}. These corrections can be applied to any given autocovariance estimate.
A general method to correct any estimator is provided by \eqref{eqn:kernel_correction} and is called as 
\begin{example}
kernel_est(estCov, kernel_name = c("gaussian", "exponential", "wave",
        "rational_quadratic", "spherical", "circular", "bessel_j", "matern", "cauchy"),
        kernel_params = c(), N_T = 0.1 * length(estCov), maxLag = length(estCov) - 1,
        x = 0:length(X), type = c("autocovariance", "autocorrelation"),
        custom_kernel = FALSE) ,
\end{example}
where \texttt{kernel\_name} and \texttt{kernel\_params} are given in Table~\ref{tab:kernel}, \texttt{N\_T} is the rate at which the kernel function vanishes at, and is recommended to be of order \(0.1 N\) when considering all lags \citep[Section~3.17]{Yaglom1987}.
As with \texttt{splines\_est}, \texttt{estCov} can be a numeric vector or a \textbf{\texttt{CovEsts}} S3 object, obtained from another autocovariance estimator.

\begin{table}[!hbp]
    \centering
    \begin{tabular}{l p{0.82\linewidth}}\toprule
        Parameter & Explanation \\ \midrule
        \texttt{x} & A vector or matrix of arguments of at least length 1. \\
        \texttt{name} & The name of the kernel. Options are: gaussian, exponential, wave, rational\_quadratic, spherical, circular, bessel\_j, matern, and cauchy. \\
        \texttt{params} & A vector of parameters for the kernel. See Table~\ref{tab:kernels} for the position of the parameters. All kernels will have a scale parameter as the first value in the vector. \\ \midrule
        Return & A vector of values. \\ \toprule
    \end{tabular}
    \caption{Example of parameters for kernel functions}
    \label{tab:kernel}
\end{table}

In addition to this, kernel-corrected versions of \eqref{eq:std_est} and \eqref{eq:std_est_pd} are provided, called by
\begin{example}
corrected_est(X, kernel_name = c("gaussian", "exponential", "wave",
        "rational_quadratic", "spherical", "circular", "bessel_j", "matern", "cauchy"),
        kernel_params = c(), N_T = 0.1 * length(X), pd = TRUE, maxLag = length(X) - 1,
        x = 0:(maxLag - 1), type = c("autocovariance", "autocorrelation"),
        meanX = mean(X), custom_kernel = FALSE) ,
\end{example}
where the parameters are the same as for \texttt{kernel\_corrected\_est} with the addition of having a positive-definite estimator. If a custom kernel is used that is not positive-definite, then the estimator will no longer be positive-definite.
For both kernel correction estimators, unlike the kernel regression estimators (recall estimators~\eqref{eq:hall_est} and \eqref{eq:hall_trunc}), the custom kernel is not required to have the properties of a symmetric probability density. 

The function that computes the taper corrected estimator~\eqref{eq:tapered_est} is called as
\begin{example}
tapered_est(X, rho, window_name = c("tukey", "triangular", "sine", "power_sine",
        "blackman", "hann_poisson", "welch"), window_params = c(1),
        maxLag = length(X) - 1, x = 0:length(X),
        type = c("autocovariance", "autocorrelation"), meanX = mean(X),
        custom_window = FALSE) ,
\end{example}
where \(\texttt{rho} \in (0, 1]\) is a scale parameter, \texttt{window\_ec} and \texttt{window\_params} are given in Table~\ref{tab:windows}, and the option \texttt{custom\_window} serves the same purpose as \texttt{custom\_kernel}. For the custom window function, it should be a nondecreasing function on \([0, 1]\) with \(w(0) = 0\) and \(w(1) = 1.\)


\subsection{Estimator adjustment/modification}
The remaining groups in Figure~\ref{fig:functions_group} fall into two categories, helper functions and \textit{Metric Functions}.
Helper functions are used during autocovariance function estimation, whilst metric functions are used to compare estimated autocovariance functions, which will be discussed in the next subsection.

The helper functions are further split into two categories, \textit{Helper Functions} and \textit{General Functions}. 
The discussion of the \textit{Helper Functions} is omitted as these functions exist only for auxiliary calculations. 
\textit{General Functions} can also be useful for other applications, for example, the forward and inverse type-II discrete cosine transforms can be called the following functions
\begin{example}
dct_1d(X) ,
idct_1d(X) ,
\end{example}
where \texttt{X} is a vector of values for which the discrete cosine transform is being computed.


Also, one can make any estimator positive-definite by using the positive-definite corrections from Section~2. 
In the package, the first correction method (see Section 2~\hyperref[make_pos_def_1]{(i)}) can be called by the function
\begin{example}
make_pd(x, method.1 = TRUE) ,
\end{example}
where \texttt{x} can be either a numeric vector of estimated autocovariance values or a \textbf{\texttt{CovEsts}} S3 object generated by one of the other autocovariance functions. If the second correction (see Section 2~\hyperref[make_pos_def_2]{(ii)}) is to be done, \texttt{method.1 = FALSE} is used instead.

Another method to make an estimated autocorrelation function positive-definite is to use linear shrinking, recall \eqref{eqn:linear_shrinking}, which can be computed using
\begin{example}
shrinking(estCov, return_matrix = FALSE, target = NULL) ,
\end{example}
where \texttt{estCov} is an estimated autocovariance or autocorrelation function, either a numeric vector or a \textbf{\texttt{CovEst}} S3 object, \texttt{return\_matrix} is a boolean, determining if the shrunken matrix to be returned or not. The default value returns a shrunken autocorrelation function instead of an autocorrelation matrix. \texttt{target} is the target matrix, which defaults to the identity matrix.

For a matrix \(\bm{A}\) that is not necessarily positive-definite, the following procedure can be used to find the nearest positive-definite matrix \cite[Th.~2.1]{Higham1988}.
First, a new matrix is constructed, \(\bm{B} = \left(\bm{A} + \bm{A}^{T} \right)\) and then compute the polar decomposition \(\bm{B} = \bm{U} \bm{H}, \) where \(\bm{U}\) is an orthogonal matrix and \(\bm{H}\) is a positive-definite matrix. The nearest positive-definite matrix to \(\bm{A}\) is then \(\bm{A}_{F} = \left(\bm{B} + \bm{H}\right) / 2 .\)
The function that computes this matrix is called using
\begin{example}
nearest_pd(X, return_matrix = FALSE) ,
\end{example}
where \code{X} is either a numeric vector, a numeric square matrix or a \textbf{\texttt{CovEsts}} S3 object. If a numeric vector or \textbf{\texttt{CovEsts}} is supplied, a matrix similar to \eqref{mat:spec_norm} will be constructed where \(D(\cdot)\) is replaced by \code{X} or the autocovariance values within the \textbf{\texttt{CovEsts}} object.

\subsection{Metrics}
Six \textit{Metric Functions} are provided in the package to compare different autocovariance estimates.

For two continuous autocovariance function estimates, \(\widehat{C}_{1}(\cdot)\) and \(\widehat{C}_{2}(\cdot)\), the area between them can be expressed as 
\(
\int_{0}^{h_{n}} \left| D(h) \right| \, \text{d}h ,
\)
where \(D(h) = \widehat{C}_{1}(h) - \widehat{C}_{2}(h)\) and \(h_{n}\) denotes the maximum lag at which the functions are estimated up to.
However, as the estimates are computed on a discrete grid, the integral is approximated using the trapezoid rule, that is,
\[
\int_{0}^{h_{n}} \left| D(h) \right| \, \text{d}h 
\approx \sum_{i = 1}^{n} \frac{\left| D(h_{i - 1}) \right| + \left| D(h_{i}) \right|}{2} (h_{i} - h_{i-1}) ,
\]
where the estimates are given over the set of lags \(\{h_{0} , h_{1} , \dots , h_{N - 1}, h_{n} \},\) and \(h_{0}\) is assumed to be 0.

The area between two estimated autocovariance functions can be calculated as 
\begin{example}
    area_between(estCov1, estCov2, lags = c(), plot = FALSE) ,
\end{example}
where \texttt{estCov1} and \texttt{estCov2} are estimated values of autocovariance functions given over the same set of lags, either numeric vectors or \textbf{\texttt{CovEsts}} S3 objects. The parameter \texttt{lags} is optional and defaults to a vector starting at 0, increasing by 1 until \texttt{length(estCov1)}. Another optional parameter, \texttt{plot}, determines whether a plot should be created showing the area between the two estimated autocovariance functions (for example, see Figure~\ref{fig:hall_area}).

The maximum distance can be expressed as follows, \( \sup_{h \in [0, h_{n}]} \left| D(h) \right| , \) where \(\sup\) is replaced by \(\max\) over the set of lags in the discrete case. 

The spectral norm is the largest eigenvalue of the matrix

\begin{equation}
\begin{bmatrix} \label{mat:spec_norm}
D(h_{0})     & D(h_{1})     & \cdots & D(h_{n - 1}) & D(h_{n})     \\
D(h_{1})     & D(h_{0})     & \cdots & D(h_{n - 2}) & D(h_{n - 1}) \\
\vdots       & \vdots       & \ddots & \vdots       & \vdots       \\
D(h_{n - 1}) & D(h_{n - 2}) & \cdots & D(h_{0})     & D(h_{1})     \\
D(h_{n})     & D(h_{n - 1}) & \cdots & D(h_{1})     & D(h_{0})     \\
\end{bmatrix}  .
\end{equation}

The maximum vertical distance and spectral norm have similar arguments to \texttt{area\_between} and are called as 
\begin{example}
max_distance(est1, est2, lags = c(), plot = FALSE) ,
spectral_norm(est1, est2) ,
\end{example}
where the plot for the maximum distance shows the magnitude of the vertical distances for each lag, see an example in Figure~\ref{fig:hall_dist}. Like with \code{area\_between}, \texttt{estCov1} and \texttt{estCov2} are estimated values of autocovariance functions given over the same set of lags, either numeric vectors or \textbf{\texttt{CovEsts}} S3 objects, which is the case for all metric functions.

The Hilbert-Schmidt metric for the matrix \eqref{mat:spec_norm} is defined as
\(
 \sqrt{\sum_{i,j= 1}^{n} d_{i,j}^{2}}
\) and can be called as
\begin{example}
hilbert_schmidt(est1, est2) .
\end{example}

The closely related mean-square error/difference (MSE) between two estimated autocovariance functions is given by
\(
n^{-1} \sum_{i = 0}^{n} D(h_{i})^{2} .
\)
If the estimate is to be compared against a theoretical model, \(\widehat{C}_{2}(\cdot)\) can be replaced with the corresponding theoretical autocovariance function.
The MSE can be called as 
\begin{example}
mse(est1, est2) .
\end{example}

To check that the autocovariance function estimate is positive-definite, first one can construct a covariance matrix that is similar to \eqref{mat:spec_norm}, where \(\widehat{C}(\cdot)\) replaces \(D(\cdot).\)
Then, if all eigenvalues of this matrix are positive, the estimate is positive-definite.
This can be checked by using the function
\begin{example}
check_pd(est) ,
\end{example}
where \texttt{est} is a vector of numeric values of an estimated autocovariance function or a \textbf{\texttt{CovEsts}} S3 object.

\section{Examples} \label{sec:examples}
This section provides an example utilisation of some main functions and computational time and memory usage analysis. It will demonstrate applications of the estimators to a simulated data set, the \texttt{sunspot.year} data set in the \pkg{datasets} package and unemployment data from the US Bureau of Labor Statistics.

The typical usage of the kernel and window functions, defined only for nonnegative values of their argument \texttt{x} (in the \texttt{window\_ec} case, it is further restricted to the interval \([0, 1]\)) is as follows:
\begin{example}
library(CovEsts)
x <- c(0.2, 0.4, 0.6)
theta <- 0.9
kernel_ec(x, "gaussian", c(theta))
[1] 0.9565287 0.8371284 0.6703200
nu <- 1
dim <- 1
kernel_ec(x, "bessel_j", c(theta, nu, dim))
[1] 0.9938398 0.9755110 0.9454638

window_ec(x, "tukey")
[1] 0.0954915 0.3454915 0.6545085
window_ec(x, "blackman", c(0.16))
[1] 0.04021286 0.20077014 0.50978714
\end{example}
When using the \texttt{bessel\_j} option, the dimension \texttt{dim} is passed, and the function additionally verifies that the kernel is valid for the specified dimension and parameters.
For the Blackman window function, one should use \(\left| a \right| \leq 0.25\) to ensure that it is nondecreasing on \([0, 1].\)

\texttt{kernel\_symm\_ec} and \texttt{window\_symm\_ec} are called in a similar way to \texttt{kernel\_ec} and \texttt{window\_ec}, but can be applied to negative values of \texttt{x}, for example,
\begin{example}
x <- c(-0.4, -0.2, 0, 0.2, 0.4)
kernel_symm_ec(x, "gaussian", c(theta))
[1] 0.4978470 0.5688553 0.5947080 0.5688553 0.4978470

window_symm_ec(x, "blackman", c(0.16))
[1] 0.7992299 0.9597871 1.0000000 0.9597871 0.7992299
\end{example}

\begin{figure}[!htb]
    \centering
    \begin{minipage}[b]{0.48\textwidth}
        \centering
        \includegraphics[width=\textwidth]{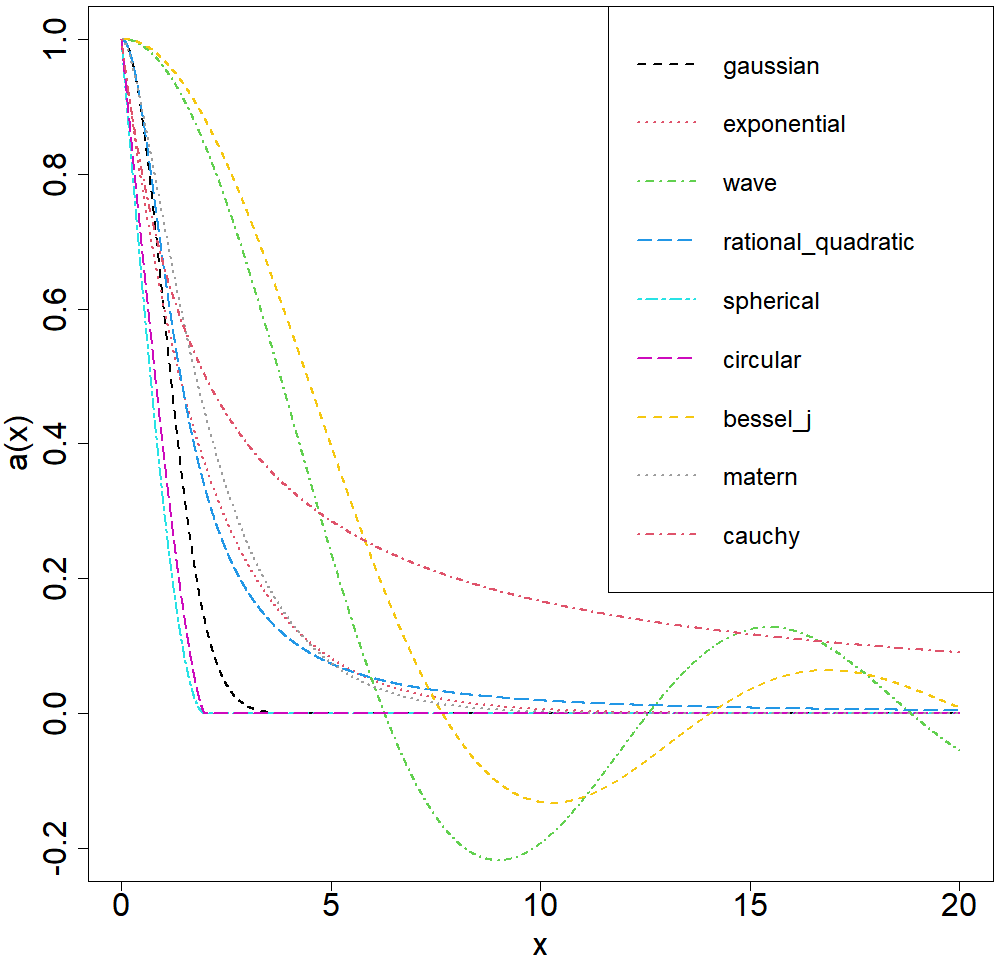}
        \caption{Examples of main available kernels where \(\theta = 2, \nu = d = \alpha = \beta = 1 .\)}
        \label{fig:kernels}
    \end{minipage}
    \hfill
    \begin{minipage}[b]{0.48\textwidth}
      \centering
        \includegraphics[width=\textwidth]{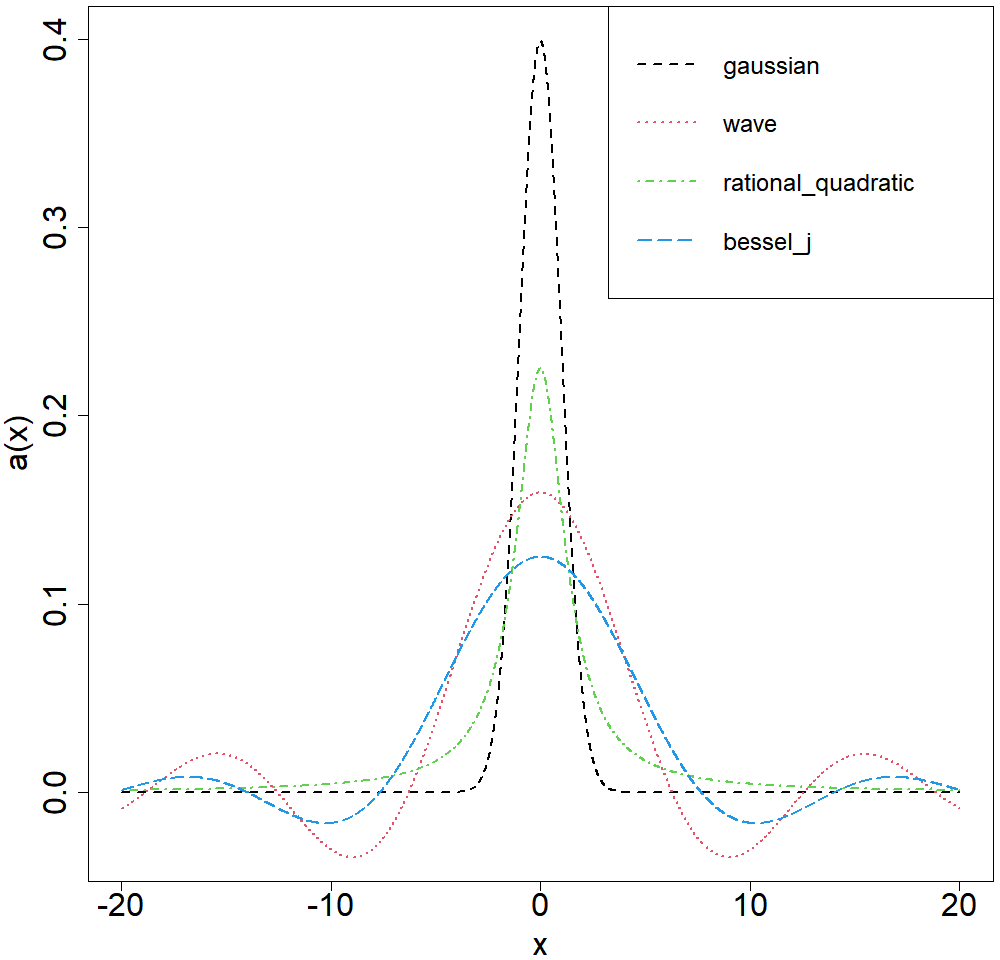}
        \caption{Examples of main available symmetric kernels where \(\theta = 2, \nu = d = \alpha = \beta = 1 .\)}
        \label{fig:kernels_symm}
    \end{minipage} 
    \end{figure}

\begin{figure}[!htb]
\centering
    \begin{minipage}[b]{0.48\textwidth}
        \centering
        \includegraphics[width=\textwidth]{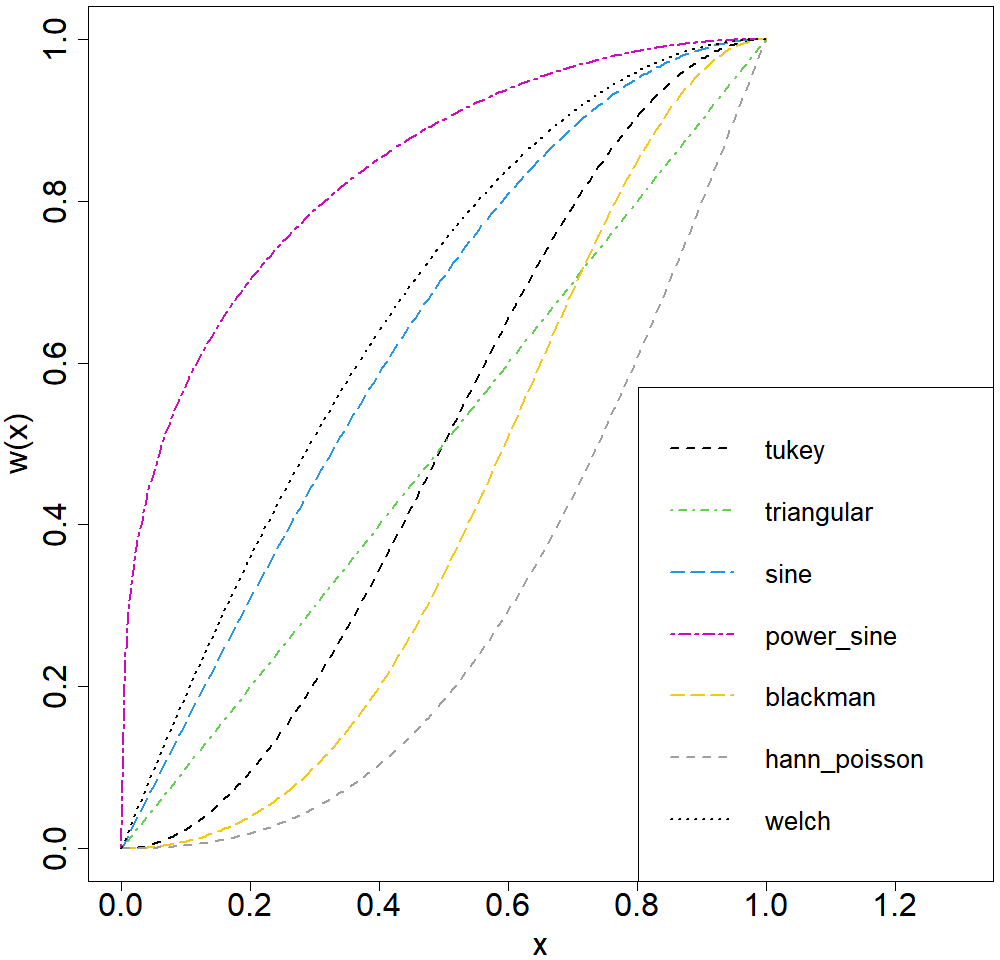}
        \caption{Examples of main available window functions where \(a = 0.3, 0.16, 1\) for the power sine, Blackman and Hann-Poisson window functions, respectively.} 
        \label{fig:windows}
    \end{minipage}
    \hfill
    \begin{minipage}[b]{0.48\textwidth}
      \centering
        \includegraphics[width=1\textwidth]{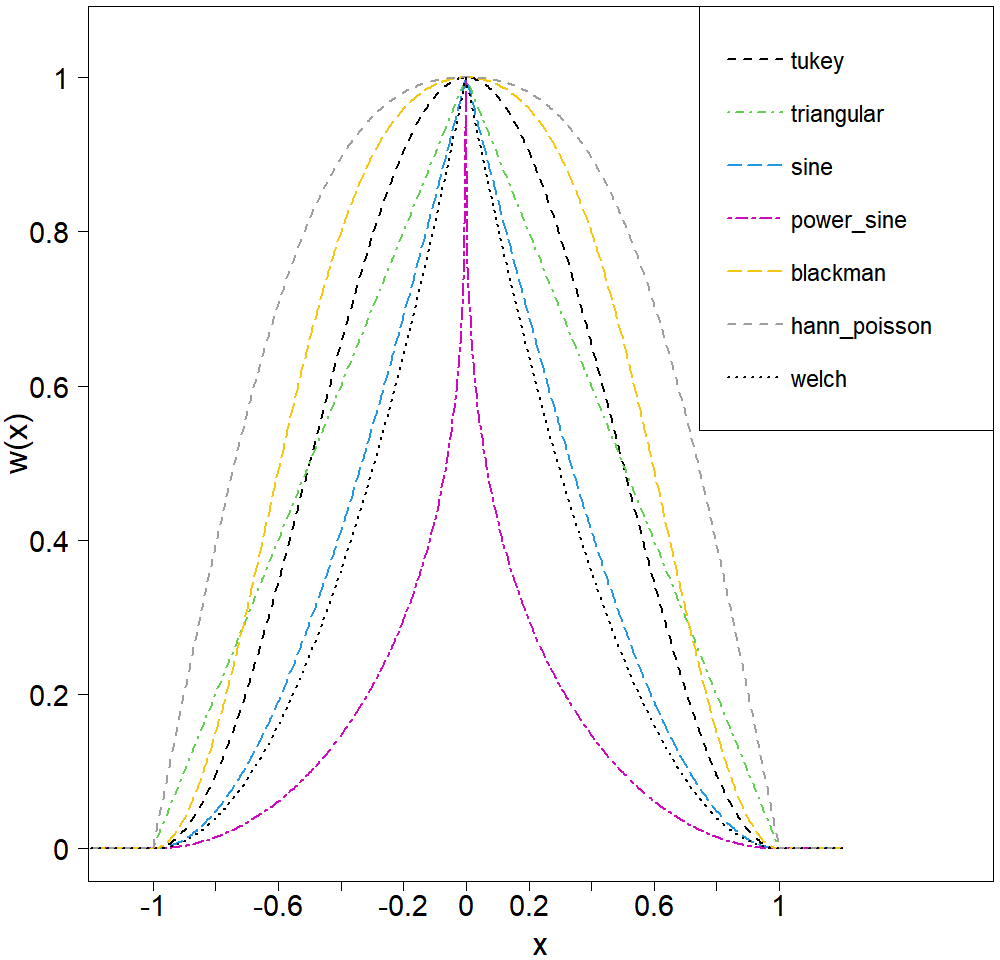}
        \caption{Examples of main available symmetric window functions where \(a = 0.3, 0.16, 1\) for the power sine, Blackman and Hann-Poisson window functions, respectively.}
        \label{fig:windows_symm}
    \end{minipage}    
\end{figure}

Figures~\ref{fig:kernels}, \ref{fig:kernels_symm}, \ref{fig:windows} and \ref{fig:windows_symm}  plot examples of the main available kernels, symmetric kernels, window functions, and symmetric window functions, respectively.

\subsection{Example 1}\label{sec:example_1}
To illustrate the application of the estimators and their accuracy, we will consider a simulated Gaussian time series on a uniform grid on \([0, 40]\) with a spacing of \(0.02,\)
having a short-range dependent Gaussian autocovariance model, \(\exp(-x^2)\).
We will only estimate the autocovariance function to a lag of 5, to use the recommended 10-20\% of total samples \citep[Section~3.17]{Yaglom1987}.

\begin{figure}[!tbp]
    \begin{minipage}[b]{0.48\textwidth}
        \centering
        \includegraphics[width=\textwidth]{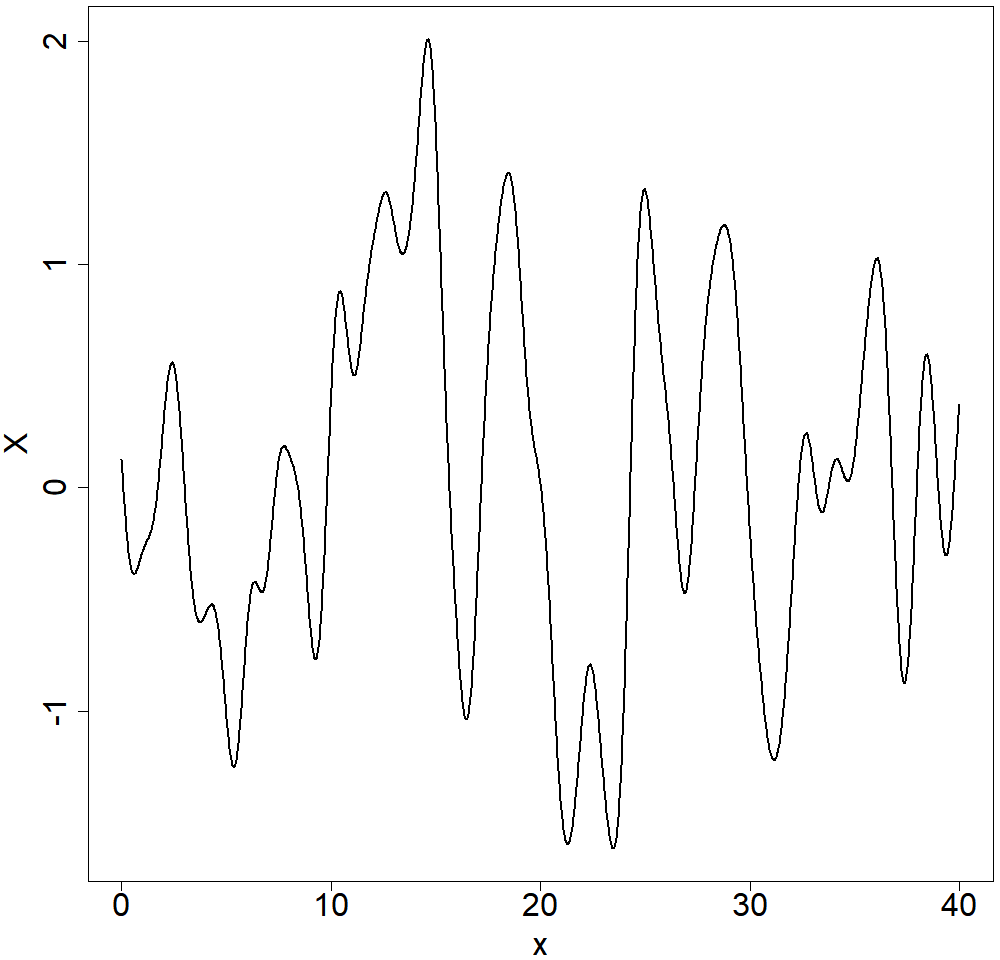}
        \caption{Realisation of short-range dependent Gaussian time series.}
        \label{fig:gaussian_realisation}
    \end{minipage}
    \hfill
    \begin{minipage}[b]{0.48\textwidth}
        \centering
        \includegraphics[width=\textwidth]{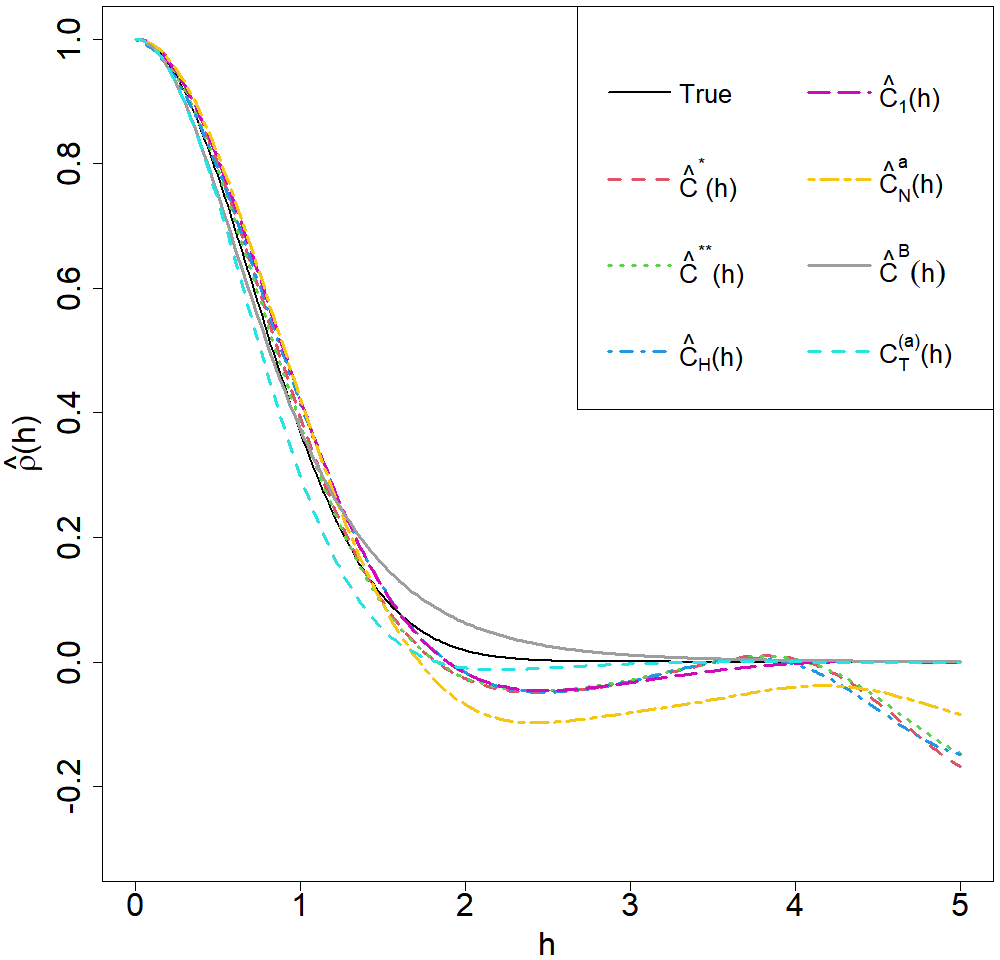}
        \caption{Estimated autocorrelation functions for a short-range dependent time series.}
        \label{fig:gaussian}
    \end{minipage}
\end{figure}

The realisations were generated using the approach based on eigendecomposition of the autocovariance matrix.
Below is the code to simulate the process and obtain the estimates based on the realisation in Figure~\ref{fig:gaussian_realisation}, which are then plotted in Figure~\ref{fig:gaussian}.
\begin{example}
set.seed(135)
N <- 2001
x <- seq(0, 40, length.out = N)
Z <- rnorm(N)
dist_mat <- abs(outer(x, x, '-'))
cov_mat <- exp(- (dist_mat^2))
eig <- eigen(cov_mat)
X <- as.vector((eig$vectors 

maxLag <- 251
t <- x[1:maxLag]

# standard estimators
Cs <- standard_est(X, maxLag = maxLag - 1, pd = FALSE, x = x)
Css <- standard_est(X, maxLag = maxLag - 1, pd = TRUE, x = x,
        type = "autocorrelation")

# Hall's estimators
hall_1 <- adjusted_est(X, x, t, 0.1, "gaussian", type = "autocorrelation")
hall_2 <- truncated_est(X, x, t, 3, 4, 0.1, "gaussian", type = "autocorrelation")

# tapered
tapered <- tapered_est(X, 1, "tukey", maxLag = maxLag - 1, x = x,
        type = "autocorrelation")

# splines
splines <- splines_est(X, x, Cs, 3, 2, maxLag = maxLag - 1, type = "autocorrelation")
Cs <- normalise_acf(Cs)

# Correction
corrected <- corrected_est(X, "gaussian", N_T=5*length(X), maxLag = maxLag - 1, x = x,
        type = "autocorrelation")

# Plot
par(mar=c(4,5.25,0.25,0.25)+.1)
plot(x[1:maxLag], exp(-x[1:maxLag]^2), type='l', lwd=2, ylim=c(-0.3, 1),
        xlab=expression(h), ylab=expression(hat(rho)*'(h)'), cex.axis=2,
        cex.lab=2)
lines(Cs, lwd=3, lty=2, col=2)
lines(Css, lwd=3, lty=3, col=3)
lines(hall_1, lwd=3, lty=4, col=4)
lines(hall_2, lwd=3, lty=5, col=6)
lines(tapered, lwd=3, lty=6, col=7)
lines(splines, lwd=3, lty=7, col=8)
lines(corrected, lwd=3, lty=8, col=13)

legend('topright', c('True', expression(hat('C')^'*'*'(h)'),
        expression(hat('C')^'**'*'(h)'), expression(hat('C')[H]*'(h)'),
        expression(hat('C')[1]*'(h)'), expression(hat('C')[N]^'a'*'(h)'),
        expression(hat('C')^'B'*('h')), expression('C'[T]^'(a)'*'(h)')),
        col=c(1, 2, 3, 4, 6, 7, 8, 13), lty=c(1, 2, 3, 4, 5, 6, 7, 8),
        lwd=c(2, rep(3, 7)), y.intersp=1, cex=2, ncol = 2)
\end{example}

Whilst for short distances, the majority of the estimated autocorrelation functions are consistent and adequately reflect the true autocorrelation function, it is clear that for increasing distances the waves start to overwhelm the estimators, seen in Figure~\ref{fig:gaussian}.
The waves are eliminated when applying the estimators \(\widehat{C}_{1}(h)\) and \(C_{T}^{(a)}(h), \) given by \eqref{eq:hall_trunc} and \eqref{eqn:kernel_correction_std_pd}, respectively. 

For this and the following examples, we consider the estimated autocorrelation functions instead of the estimated autocovariance functions to allow for easier comparison.

\begin{figure}[!htb]
    \begin{minipage}[b]{0.48\textwidth}
        \centering
        \includegraphics[width=\textwidth]{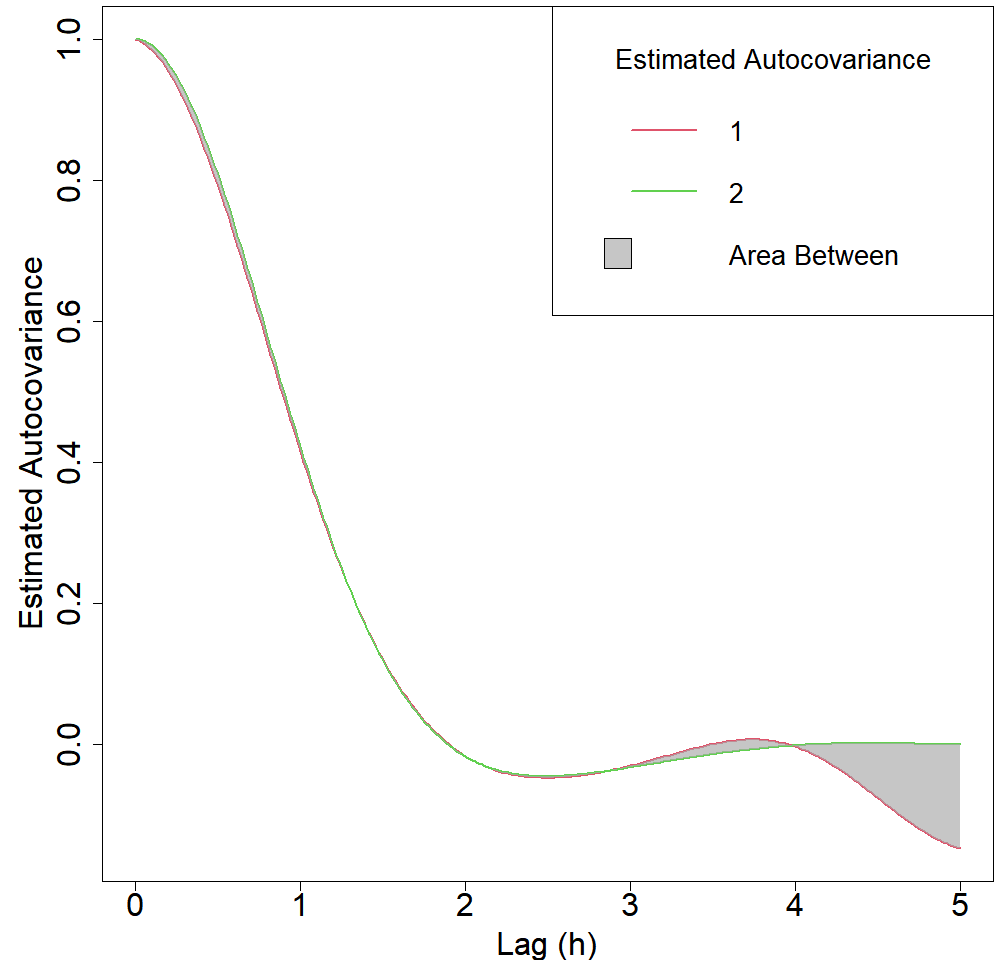}
        \caption{Area between estimated autocovariance functions.}
        \label{fig:hall_area}
    \end{minipage}
    \hfill
    \begin{minipage}[b]{0.48\textwidth}
        \centering
        \includegraphics[width=\textwidth]{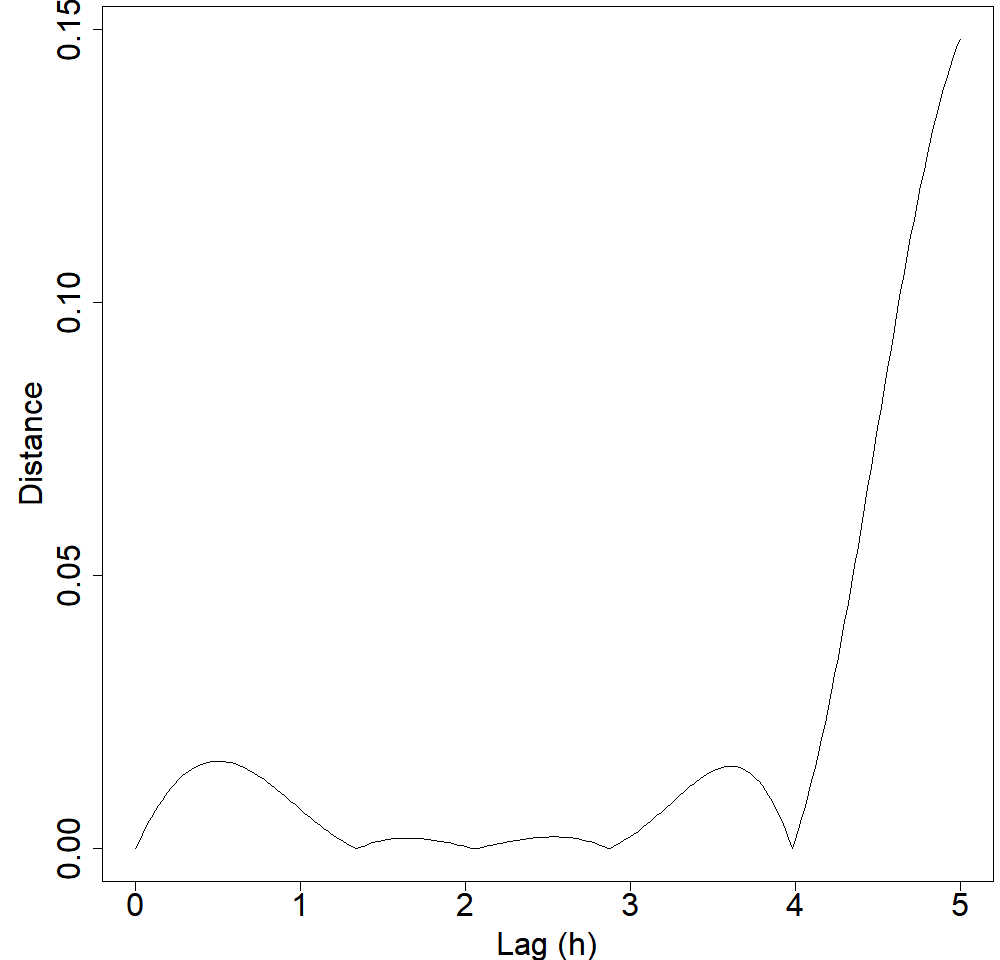}
        \caption{Distances between estimated autocovariance functions.}
        \label{fig:hall_dist}
    \end{minipage}
\end{figure}

Figure~\ref{fig:hall_area} plots the area (in grey) between estimators given by~\eqref{eq:hall_est} (red) and \eqref{eq:hall_trunc} (green). As the green curve is forced to zero after \(h=4,\) the area between the two estimates increases after that point. The estimated area between them is approximately 0.061.
This difference can also be seen in the plot of distances between the two estimates in Figure~\ref{fig:hall_dist}. For \(h < 3,\) the estimates are rather close. However, the distance starts to increase after this point dramatically. The maximum distance between the two estimates is 0.1.
These values and plots can be produced as follows:
\begin{example}
area_between(hall_1, hall_2, plot=T)
max_distance(hall_1, hall_2, plot=T)
\end{example}

\begin{figure}[!htb]
    \begin{minipage}[b]{0.48\textwidth}
        \centering
        \includegraphics[width=\textwidth]{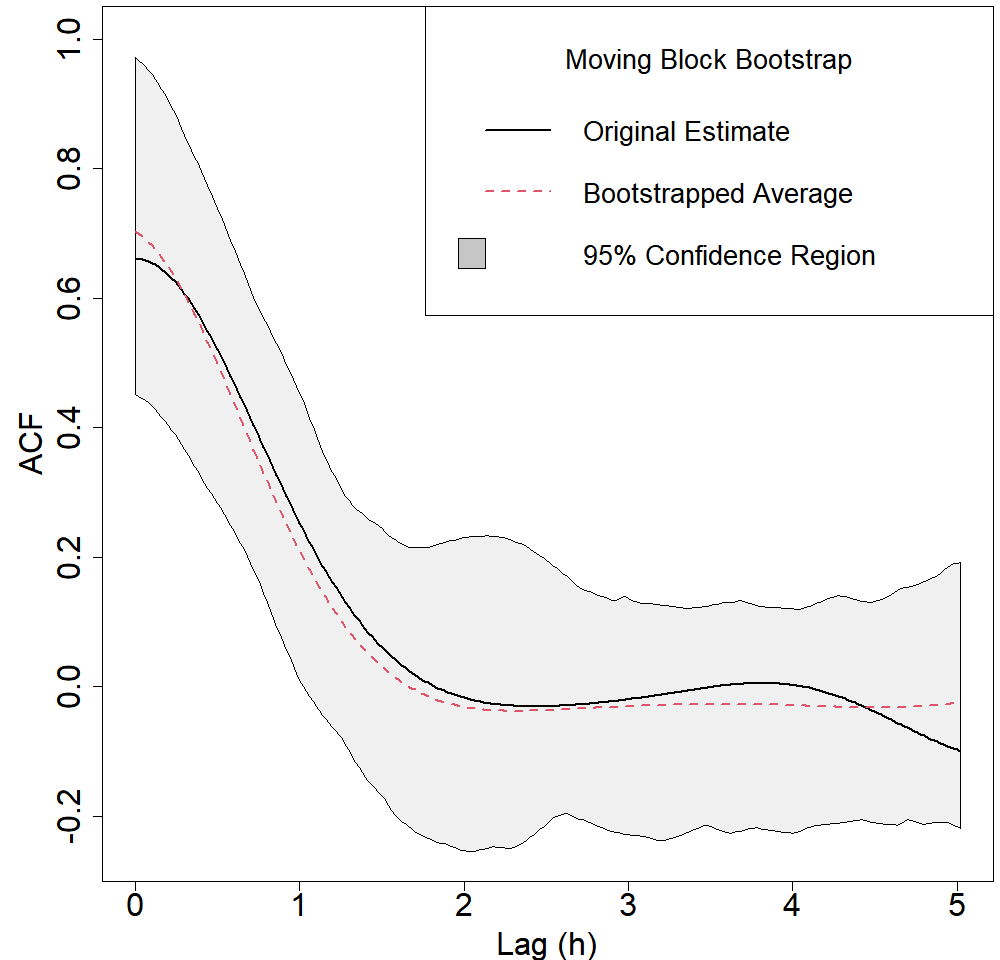}
        \caption{Moving block bootstrap estimates and  confidence region.}
        \label{fig:moving_boot}
    \end{minipage}
    \hfill
    \begin{minipage}[b]{0.48\textwidth}
        \centering
        \includegraphics[width=\textwidth]{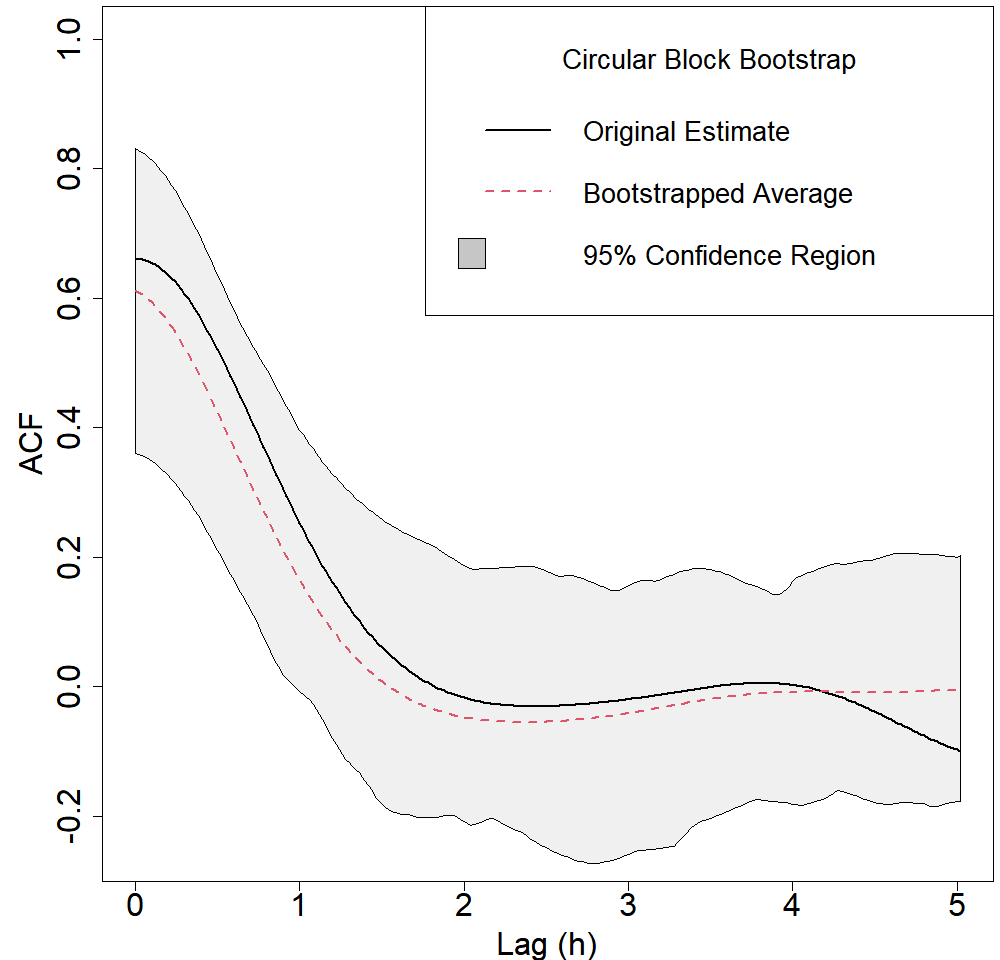}
        \caption{Circular block bootstrap estimates and  confidence region.}
        \label{fig:circular_boot}
    \end{minipage}
\end{figure}

To compute the moving block and circular bootstrap for estimator~\eqref{eq:std_est_pd}, one can use the following commands
\begin{example}
plot(block_bootstrap(X, maxLag, x, l = maxLag), ylim=c(-0.25, 1), cex.axis=2, cex.lab=2)
plot(block_bootstrap(X, maxLag, x, l = maxLag, boot_type = 'circular'),
        ylim=c(-0.25, 1), cex.axis=2, cex.lab=2)
\end{example}
Figures~\ref{fig:moving_boot} and \ref{fig:circular_boot} show graphical outputs from the \code{block\_boostrap} function, where the autocovariance functions are plotted. The solid black line is the estimated autocovariance function, the red dashed line is the average bootstrap autocovariance function, and the grey shaded bootstrap 95\% confidence region is calculated pointwise for each lag. Both bootstrap estimates give similar results, with the bootstrap estimate lessening constant summation effects, compared to the original estimate that dips after lag 4.

\subsection{Example 2}
\begin{figure}[!htb]
    \centering
    \begin{minipage}[b]{0.48\textwidth}
        \centering
        \includegraphics[width=\textwidth]{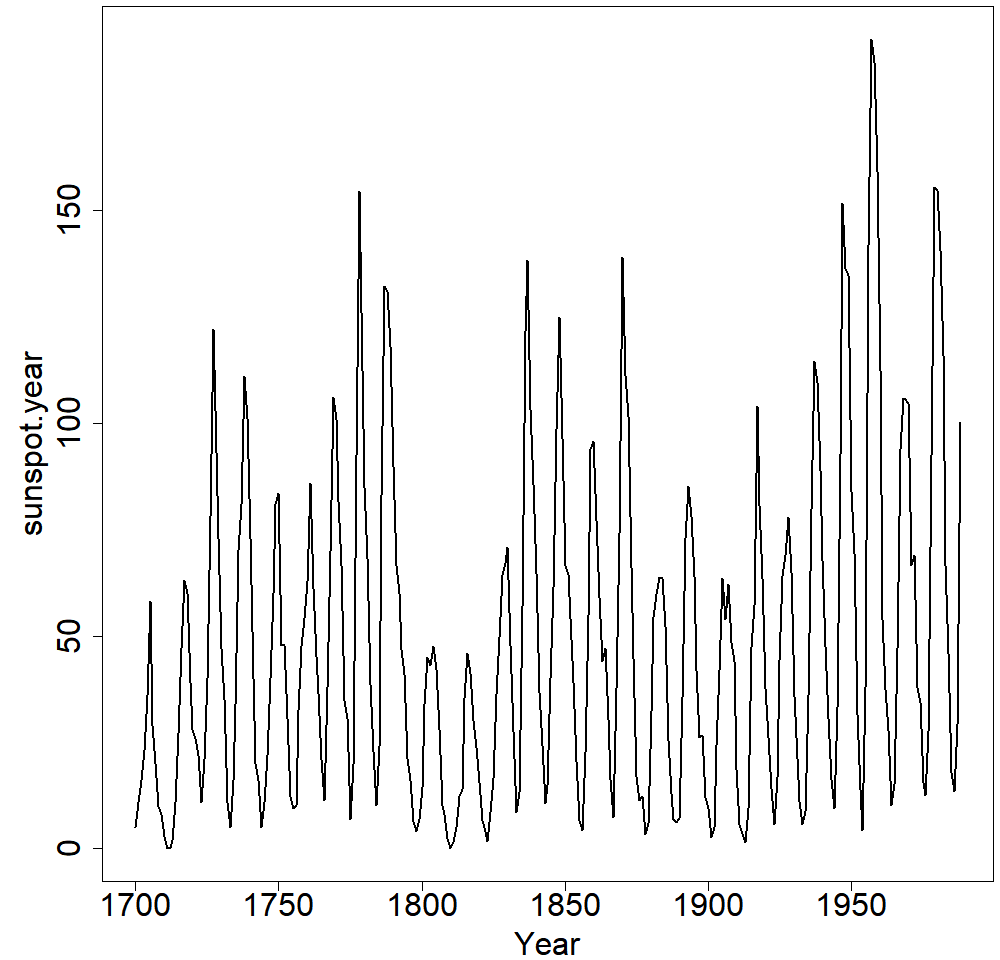}
        \caption{Yearly sunspots count data, from 1700 to 1988.}
        \label{fig:sunspots_data}
    \end{minipage}
    \hfill
        \centering
    \begin{minipage}[b]{0.48\textwidth}
        \centering
        \includegraphics[width=\textwidth]{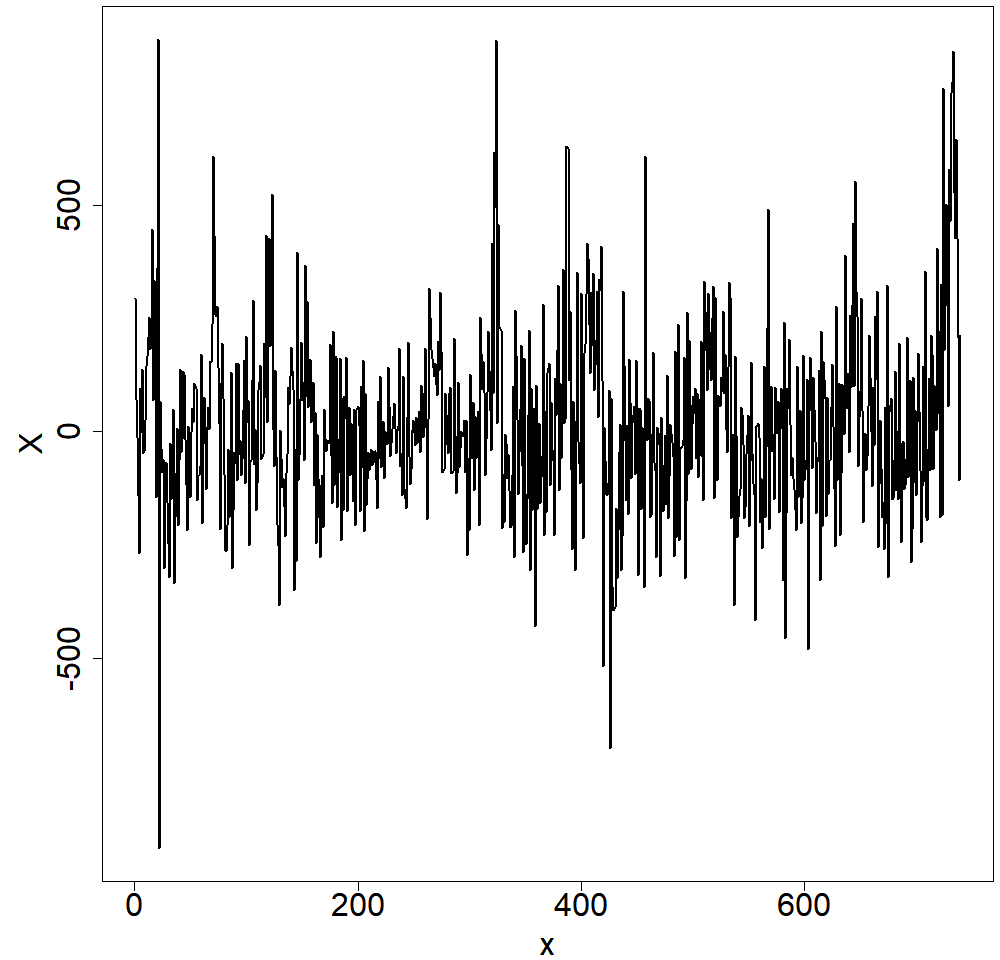}
        \caption{Increments of US unemployment count data.}
        \label{fig:bls_series}
    \end{minipage}
    \end{figure}

\begin{figure}[!htb]
    \centering    
    \begin{minipage}[b]{0.48\textwidth}
      \centering
        \includegraphics[width=\textwidth]{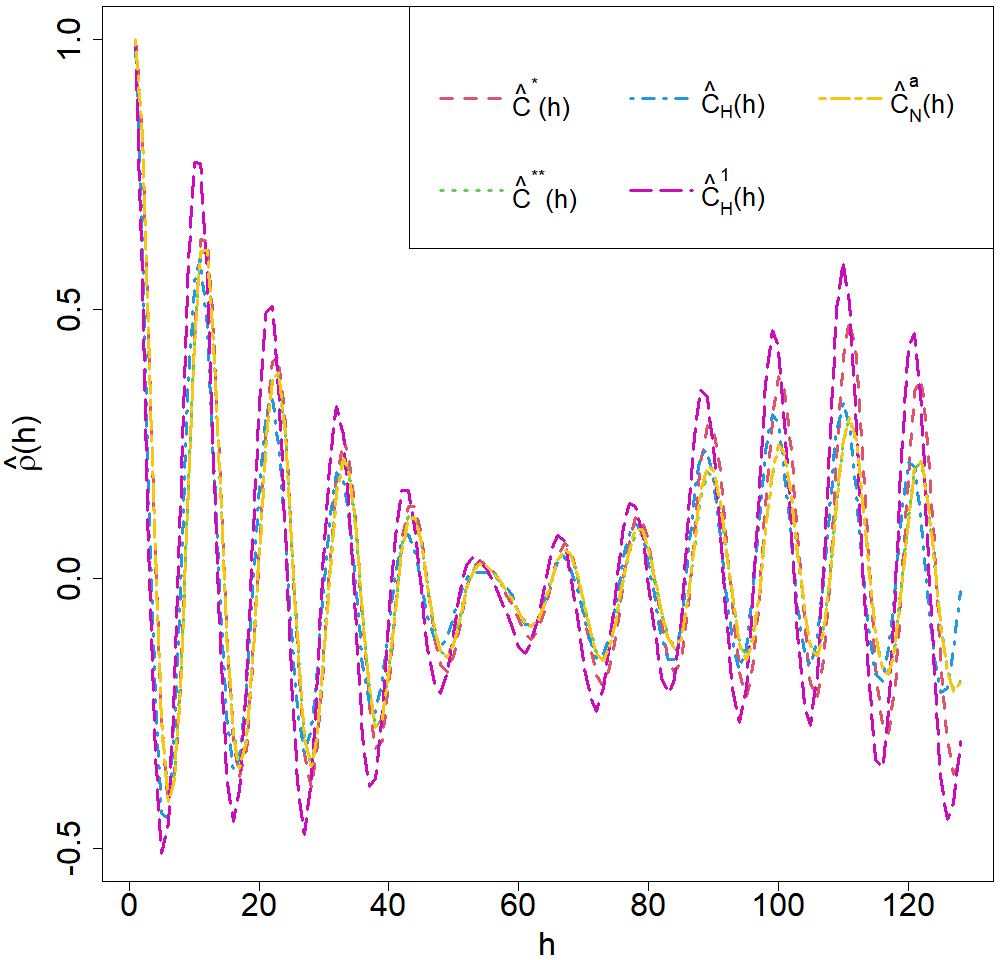}
        \caption{Estimated autocorrelation functions for the sunspots data.}
        \label{fig:sunspots}
    \end{minipage}   \hfill
        \centering
          \begin{minipage}[b]{0.48\textwidth}
      \centering
    \includegraphics[width=\textwidth]{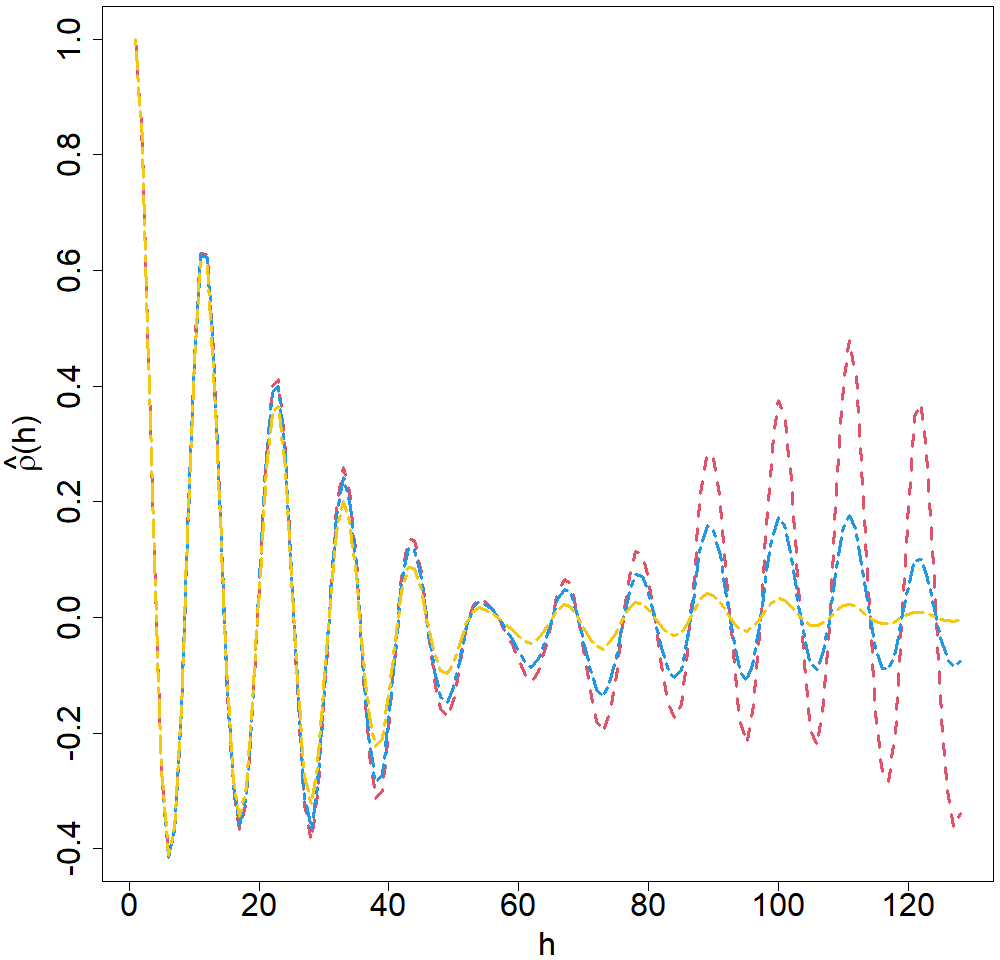}
    \caption{Smoothing of \(C^{*}(h)\) autocorrelation estimate.} 
    \label{fig:smooth_sunspot}
      \end{minipage}
\end{figure}

This example uses the classical yearly sunspot count data from 1700 to 1988, \texttt{sunspot.year}, available in the \pkg{datasets} R package. This data gives the number of observed yearly sunspots, seen in Figure~\ref{fig:sunspots_data}.
It has been shown that the number of sunspots exhibits long-range dependent and cyclic behaviours, see \citet{GilAlana2009} and \citet{Hu2009}, with an approximate 11-year cycle. So, one should expect to see periodicity in the autocovariance estimate at multiples of lag 11.
The example demonstrates that estimator~\eqref{eq:hall_trunc} has potential issues as its Fourier transform at the corresponding second nonzero frequency takes a negative value. So, only one nonzero frequency can be used to obtain the corrected estimates. Thus, this positive-definite adjustment will not produce any useful result and is not reported.  Instead, it is replaced with the non-positive-definite version of estimator~\eqref{eq:hall_est}, denoted by \(\widehat{C}_{H}^{1}(h)\) in the legend of Figure~\ref{fig:sunspots}. The splines estimator will also be omitted as it does not capture the cyclicality in the autocovariance function. 
\begin{example}
X <- as.vector(sunspot.year)
x <- 1:length(X)
maxLag <- 128

# standard estimators
Cs <- standard_est(X, maxLag = maxLag - 1, pd = FALSE, meanX = mean(X), x = x,
        type = "autocorrelation")
Css <- standard_est(X, maxLag = maxLag - 1, pd = TRUE, meanX = mean(X), x = x,
        type = "autocorrelation")

# Hall's estimators
hall_1 <- adjusted_est(X, x, x[1:maxLag], b = 0.1, kernel_name = "wave",
        type = "autocorrelation")
hall_2 <- adjusted_est(X, x, x[1:maxLag], b = 0.1, kernel_name = "wave", pd = FALSE,
        type = "autocorrelation")

# tapered
tapered <- tapered_est(X, 0.01, "tukey", maxLag = maxLag - 1, x = x,
        type = "autocorrelation")

par(mar=c(4,5.25,0.25,0.25)+.1)

plot(Cs, lwd=3, lty=2, col=2, type='l', ylim=c(-0.5, 1), xlab=expression(h),
        ylab=expression(hat(rho)*'(h)'), cex.axis=2, cex.lab=2)
lines(Css, lwd=3, lty=3, col=3)
lines(hall_1, lwd=3, lty=4, col=4)
lines(hall_2, lwd=3, lty=5, col=6)
lines(tapered, lwd=3, lty=6, col=7)

legend('topright', c(expression(hat('C')^'*'*'(h)'), expression(hat('C')^'**'*'(h)'),
        expression(hat('C')[H]*'(h)'), expression(hat('C')[H]^'1'*'(h)'),
        expression(hat('C')[N]^'a'*'(h)')), col=c(2, 3, 4, 6, 7),
        lty=c(2, 3, 4, 5, 6), lwd=c(rep(3, 5)), y.intersp=1, cex=1.7, ncol=3)
\end{example}

As shown in Figure~\ref{fig:sunspots}, the estimators provide consistent results. However, since no seasonal adjustment or transformation was applied, the estimates may not correspond to the optimal model specification for these data.


We have provided an example of how waves can be reduced and eliminated from the estimates in Figure~\ref{fig:smooth_sunspot}. The red line, the original estimate (computed using estimator~\eqref{eq:std_est}), behaves cyclically and has waves of larger amplitude as the estimation lag increases. The blue and yellow lines show the smoothing results by using the wave (with \(\theta = 50)\)) and Gaussian (with \(\theta = 4000\)) kernels, respectively. Such kernels were chosen to lessen constant summation and other unwanted effects and to bring the estimator to zero. The latter correction may be appropriate when short-range dependence is expected.

\subsection{Example 3}
We use monthly changes in U.S. unemployment counts data studied by \citet{Artiach2011} and sourced from the U.S. Bureau of Labor Statistics \url{https://www.bls.gov/}. There are 739 months of observations in the time series plotted in Figure~\ref{fig:bls_series}.
We estimated the autocovariance function up to a lag of 144, representing a maximum difference in 12 years between observations.  In \cite{Artiach2011}, the periodogram of this time series has a nonzero singularity indicating cyclic long-memory, see~\cite{Olenko2022}, where each cycle is roughly 5.5 years. So, we should expect some cyclicality in the estimated autocovariance functions.
\begin{figure}[!htb]
\centering
    \begin{minipage}[b]{0.48\textwidth}
      \centering
        \includegraphics[width=\textwidth]{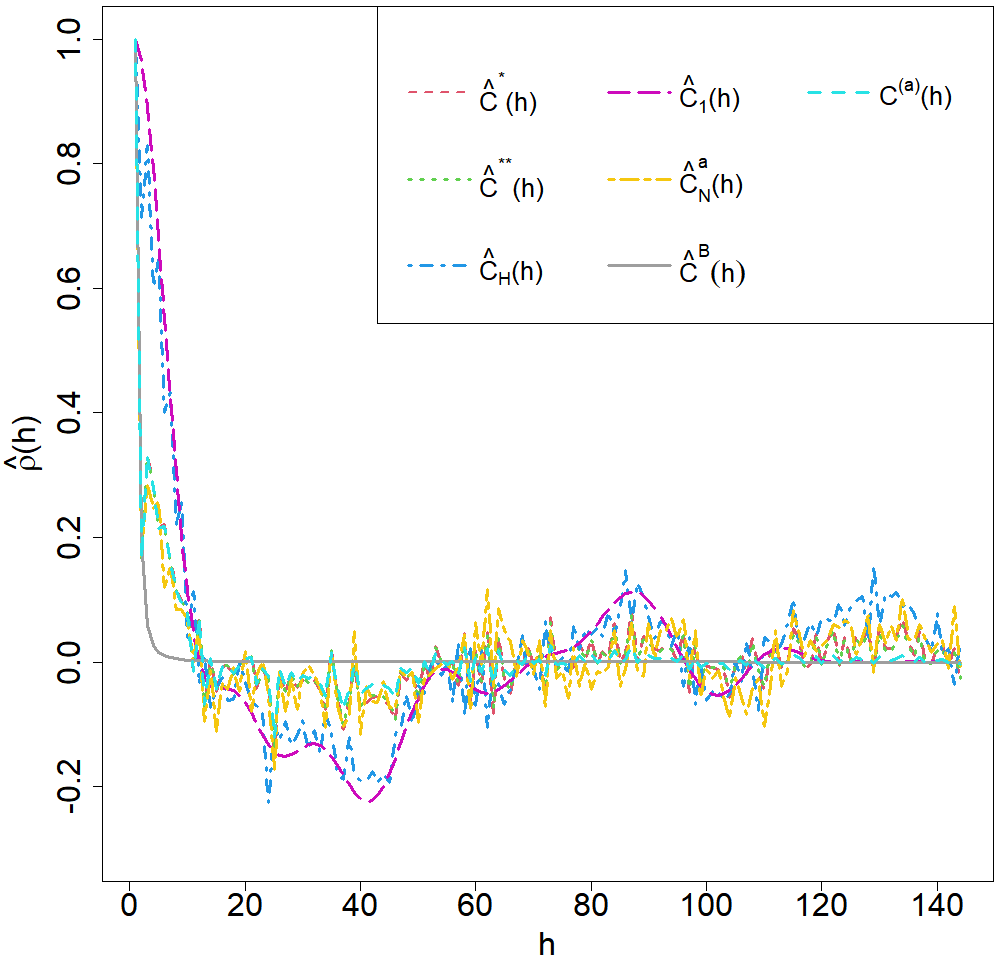}
        \caption{Estimated autocorrelation functions for unemployment increments data.}
        \label{fig:bls}
    \end{minipage} \hfill
     \begin{minipage}[b]{0.48\textwidth}
      \centering
    \includegraphics[width=\textwidth]{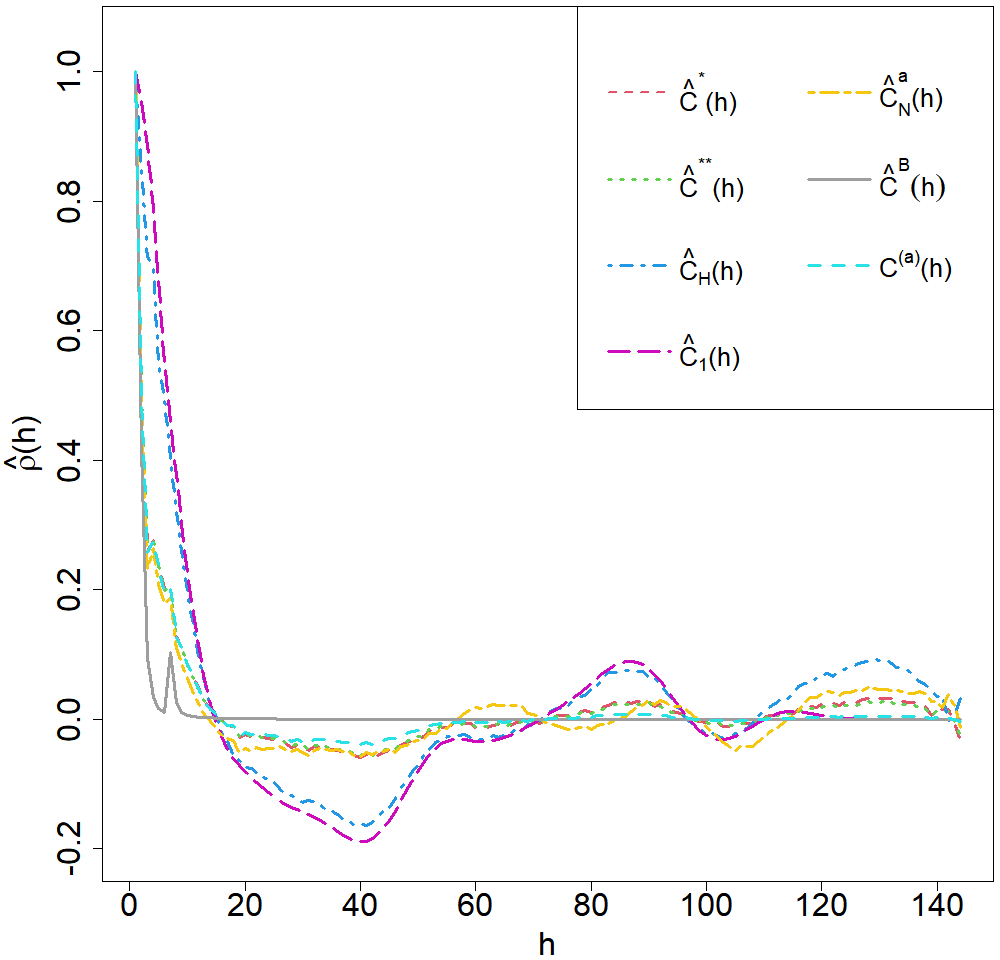}
    \caption{Smoothing of estimates for the unemployment increments data.} 
    \label{fig:smooth_bls}
      \end{minipage}
\end{figure}

The following code produces the estimated autocorrelations plotted in Figure~\ref{fig:bls}.
\begin{example}
maxLag <- 144
X <- X_bls
x <- 1:length(X)

# standard estimators
Cs <- standard_est(X, maxLag = maxLag - 1, pd = FALSE, x = x)
Css <- standard_est(X, maxLag = maxLag - 1, pd = TRUE, x = x,
        type = "autocorrelation")

# Hall's estimators
hall_1 <- adjusted_est(X, x, x[1:maxLag], b = 0.1, kernel_name = "rational_quadratic",
        type = "autocorrelation")
hall_2 <- truncated_est(X, x, x[1:maxLag], 110, 120, b = 0.1,
        kernel_name = "rational_quadratic", type = "autocorrelation")

# tapered
tapered <- tapered_est(X, 1, "tukey", maxLag = maxLag - 1, x = x,
        type = "autocorrelation")

# splines
splines <- splines_est(X, x, Cs, 3, 2, maxLag = maxLag - 1, type = "autocorrelation")
Cs <- normalise_acf(Cs)

# Correction
corrected <- corrected_est(X, "rational_quadratic", N_T = 5*length(X), maxLag = maxLag - 1,
        x = x, type = "autocorrelation")

colours <- c(2, 3, 4, 6, 7, 8, 13)
ltys <- c(2, 3, 4, 5, 6, 7, 8)
par(mar=c(4,5.25,0.25,0.25)+.1)
plot(Cs, lwd=3, lty=2, col=2, type='l', ylim=c(-0.3, 1), xlab=expression(h),
        ylab=expression(hat(rho)*'(h)'), cex.axis=2, cex.lab=2)
lines(Css, lwd=3, lty=3, col=3)
lines(hall_1, lwd=3, lty=4, col=4)
lines(hall_2, lwd=3, lty=5, col=6)
lines(tapered, lwd=3, lty=6, col=7)
lines(splines, lwd=3, lty=7, col=8)
lines(corrected, lwd=3, lty=8, col=13)

legend('topright', c(expression(hat('C')^'*'*'(h)'), expression(hat('C')^'**'*'(h)'),
        expression(hat('C')[H]*'(h)'), expression(hat('C')[1]*'(h)'),
        expression(hat('C')[N]^'a'*'(h)'), expression(hat('C')^'B'*('h')),
        expression('C'^'(a)'*'(h)')), col=colours, lty=ltys, lwd=c(2, rep(3, 6)),
        y.intersp=1.2, cex=1.8, ncol=3)
\end{example}

As can be seen from Figure~\ref{fig:bls}, the estimates have local chaotic fluctuations. To better see the main pattern of the estimates, Figure~\ref{fig:smooth_bls} shows the application of a moving average to each of the estimators in Figure~\ref{fig:bls}. After this smoothing, two estimators, \eqref{eq:hall_est} and \eqref{eq:hall_trunc}, exhibit cyclicality expected for this data.

\subsection{Example 4}
This example empirically investigates the computational complexity of the methods by evaluating the runtime and memory usage of the estimators. Autocovariance function estimates were computed for a sequence of simulated time series with Gaussian autocovariance, similar to \hyperref[sec:example_1]{Example 1}. The realisations of the time series were simulated on uniform grids over \([0, 40]\) with lengths \(N=101, 151, \dots, 951, 1001\).
This process was repeated 100 times, and the results presented are the medians.

\begin{figure}[!htbp]
\centering
    \begin{minipage}[b]{0.48\textwidth}
      \centering
        \includegraphics[width=\textwidth]{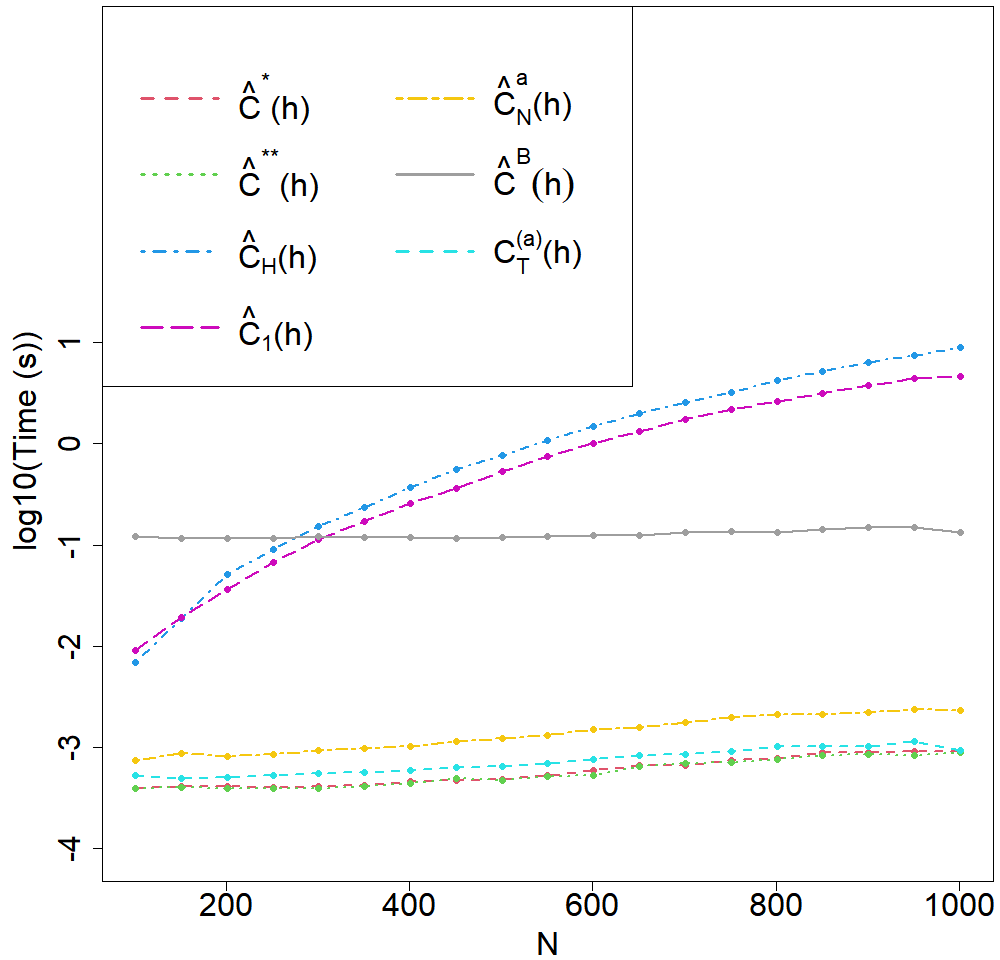}
        \caption{Log-scale computational time (s) for estimates for time series of varying length.}
        \label{fig:time_all}
    \end{minipage} \hfill
     \begin{minipage}[b]{0.48\textwidth}
      \centering
    \includegraphics[width=\textwidth]{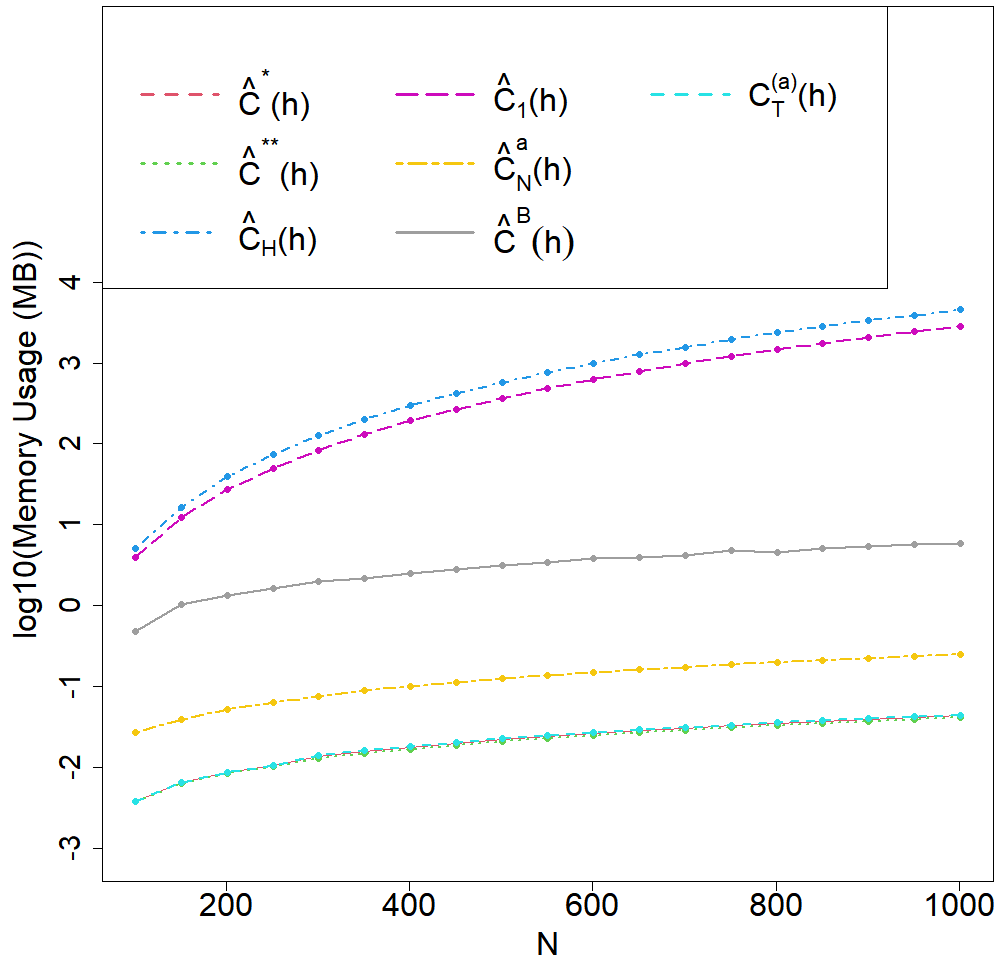}
    \caption{Log-scale memory usage (MB) for estimates for time series of varying length.} 
    \label{fig:memory_all}
      \end{minipage}
\end{figure}

Figure~\ref{fig:time_all} shows the log10-time of the computed estimators. The log scale was used due the large difference between estimators. For example, estimator~\eqref{eq:std_est_pd} was computed in 0.000781015  seconds, on average, for a time series of length \(N = 1001,\) whilst estimator~\eqref{eq:hall_est} was computed in 8.45674 seconds.
It is clear that all estimators~\eqref{eq:std_est}, \eqref{eq:std_est_pd}, \eqref{eq:tapered_est}, and \eqref{eqn:kernel_correction_std_pd} perform similarly in terms of time, whilst Hall's estimators~\eqref{eq:hall_est} and \eqref{eq:hall_trunc} exhibit significant growth in computational time. The splines estimator~\eqref{eq:splines_est} shows a consistent time usage across all estimates, exhibiting minimal growth as \(N\) increases. 

Figure~\ref{fig:memory_all} shows the log10-memory usage, where the log-scale was chosen for the same reasons as the time case.
The estimators~\eqref{eq:std_est}, \eqref{eq:std_est_pd}, and \eqref{eqn:kernel_correction_std_pd} all effectively used the same memory. The estimator~\eqref{eq:tapered_est} is quite close, particularly in terms of growth; however, it is shifted upward. Again, Hall's estimators~\eqref{eq:hall_est} and \eqref{eq:hall_trunc} exhibit significantly greater memory usage growth than the other estimators. The splines estimator~\eqref{eq:splines_est} is stable in terms of growth, as in the time case.

It is clear that Hall's estimators, \eqref{eq:hall_est} and \eqref{eq:hall_trunc}, require significantly more computational time and memory usage compared to other estimators. Thus, these estimators may not be appropriate in the case of Big Data. The other estimators are computed relatively quickly and require significantly less memory, even for large values of \(N.\) Further, these limitations should be considered when performing block bootstrap or other statistical procedures involving repeated computation of estimated covariances (e.g., moving window analysis, model diagnostics, Kriging/Gaussian process prediction and interpolation), as the estimator must be called many times. In such cases, using parallelism options, as implemented in the package for bootstrap, may substantially reduce computational time.

\section{Summary}
The article introduced the package \CRANpkg{CovEsts}, which provides several nonparametric autocovariance function estimators available in the literature. The package offers a set of functions that allow users to easily apply the different estimators, as shown in the examples. The autocovariance estimator functions have a high level of flexibility with various options that the user can control, such as a custom kernel, selecting the sampling grid, and controlling the positive-definiteness of estimators, among others. Several functions are provided to generate bootstrap autocovariance confidence regions and to address potential issues in autocovariance estimation. In addition, various metrics are included to compare the performance of the estimators.
In the future, the package is planned to be extended to work with similar nonparametric estimators for spatial data.

\section{Computation details}
R version 4.5 was used to develop, test the package, and produce all outputs in this article.

\section{Acknowledgement}
This research was partially supported by the Australian Research Council Discovery Projects funding scheme (project DP220101680). Andriy Olenko was also partially supported by La Trobe University's SCEMS CaRE and Beyond grant. The authors are grateful to Professors N. Cressie, N. Leonenko, and E. Porcu for their feedback on certain approaches to the nonparametric estimation of covariance functions.
We also thank the editor, Prof. R.~J.~Hyndman, and anonymous reviewers for their comments that helped to improve the package and the paper.

\bibliography{bilchouris-olenko}

\address{Adam Bilchouris\\
  Department of Mathematics and Statistics\\
  La Trobe University\\
  Australia\\
  ORCiD: 0009-0002-4649-0247\\
  \email{a.bilchouris@latrobe.edu.au}}

\address{Andriy Olenko\\
  Department of Mathematics and Statistics\\
  La Trobe University\\
  Australia\\
  ORCiD: 0000-0002-0917-7000\\
  \email{a.olenko@latrobe.edu.au}}

\end{article}

\end{document}